\documentclass[useAMS,usenatbib]{mn2e}
\usepackage{subfigure}
\usepackage{graphicx}

\title[Spectropolarimetry of Seyfert 2 galaxies]{Upholding the Unified 
Model for Active Galactic Nuclei: VLT/FORS2 Spectropolarimetry of Seyfert 2 galaxies}
\author[C. Ramos Almeida et al.]
{\parbox{\textwidth}{C. Ramos Almeida$^{1,2}$\thanks{Ram\'on y Cajal Fellow. E-mail: cra@iac.es},
M. J. Mart\' inez Gonz\'alez$^{1,2}$, 
A. Asensio Ramos$^{1,2}$, 
J. A. Acosta-Pulido$^{1,2}$, 
S. F. H\"onig$^{3}$, 
A. Alonso-Herrero$^{4,5}$, 
C. N. Tadhunter$^{6}$
O. Gonz\'alez-Mart\' in$^{7}$
}\vspace{0.4cm}\\
\parbox{\textwidth}{$^{1}$Instituto de Astrof\' isica de Canarias, Calle V\' ia L\'actea, s/n, E-38205, La Laguna, Tenerife, Spain\\
$^{2}$Departamento de Astrof\' isica, Universidad de La Laguna, E-38205, La Laguna, Tenerife, Spain\\
$^{3}$School of Physics \& Astronomy, University of Southampton, Southampton, SO17 1BJ, UK\\
$^{4}$Centro de Astrobiolog\'{\i}a (CAB, CSIC-INTA), ESAC Campus, E-28692, Villanueva de la Ca\~nada, Madrid, Spain\\
$^{5}$Department of Physics and Astronomy, University of Texas at San Antonio, One UTSA Circle, San Antonio, TX 78249, USA\\ 
$^{6}$Department of Physics \& Astronomy, University of Sheffield, Sheffield S3 7RH, UK\\
$^{7}$Instituto de Radioastronom\' ia y Astrof\' isica (IRAF-UNAM), 3-72 (Xangari), 8701, Morelia, Mexico
}}

\begin{document}

\date{}

\pagerange{\pageref{firstpage}--\pageref{lastpage}} \pubyear{2015}

\maketitle

\label{firstpage}

%\author{C. Ramos Almeida\altaffilmark{1,2,\dagger}, M. J. Mart\' inez Gonz\'alez\altaffilmark{1,2}, A. Asensio Ramos\altaffilmark{1,2}, 
%J. A. Acosta-Pulido\altaffilmark{1,2}, S. F. H\"onig\altaffilmark{3,\dagger}, A. Alonso-Herrero\altaffilmark{4,5}, 
%C. N. Tadhunter\altaffilmark{6}}
%\email{cra@iac.es}


%\altaffiltext{1}{Instituto de Astrof\'\i sica de Canarias, C/V\'\i a L\'{a}ctea, s/n, E-38205, La Laguna, Tenerife, Spain}
%\altaffiltext{2}{Departamento de Astrof\' isica, Universidad de La Laguna, E-38205, La Laguna, Tenerife, Spain}
%\altaffiltext{3}{School of Physics \& Astronomy, University of Southampton, Southampton, SO17 1BJ, UK}
%\altaffiltext{4}{Instituto de F\' isica de Cantabria, CSIC-Universidad de Cantabria, E-39005, Santander, Spain}
%\altaffiltext{5}{Department of Physics and Astronomy, University of Texas at San Antonio, One UTSA Circle, San Antonio, TX 78249, USA}
%\altaffiltext{6}{Department of Physics \& Astronomy, University of Sheffield, Sheffield S3 7RH, UK}
%\altaffiltext{$\dagger$}{Marie Curie Fellow.}


\begin{abstract}
The origin of the unification model for Active Galactic Nuclei (AGN) was the detection of broad hydrogen recombination
lines in the optical polarized spectrum of the Seyfert 2 galaxy (Sy2) NGC\,1068. Since then, a search for the 
hidden broad-line region (HBLR) of nearby Sy2s started, but polarized broad lines have only been detected in 
$\sim$30--40\% of the nearby Sy2s observed to date. Here we present
new VLT/FORS2 optical spectropolarimetry of a sample of 15 Sy2s, including Compton-thin and Compton-thick sources. 
The sample includes six galaxies without previously published spectropolarimetry, 
some of them normally treated as non-hidden BLR (NHBLR) objects in the literature, 
four classified as NHBLR, and five as HBLR based on previous data.
We report $\ge$4$\sigma$ detections of a HBLR in 11 of these galaxies (73\% of the sample) and a tentative 
detection in NGC\,5793, 
which is Compton-thick according to the analysis of X-ray data performed here. 
Our results confirm that at least some NHBLRs are misclassified, bringing previous publications reporting 
differences between HBLR and NHBLR objects into question. We detect broad H$\alpha$ and H$\beta$ components in polarized 
light for 10 targets, and just broad H$\alpha$ for NGC\,5793 and NGC\,6300, with line widths ranging 
between 2100 and 9600 km~s$^{-1}$. High bolometric luminosities and low column densities are associated with higher
polarization degrees, but not necessarily with the detection of the scattered broad components.
%We do not find any correlation between the properties of the polarized spectra and galaxy inclination or torus opening angle and inclination, 
%but a larger sample is required to confirm this lack of correlation. 
\end{abstract}


%\keywords{galaxies: active --- galaxies: nuclei --- galaxies: Seyfert --- polarization}



\section{Introduction}

The unified model for active galactic nuclei (AGN; \citealt{Antonucci93}) stems from the detection of 
permitted broad lines in the optical polarized spectrum of the Seyfert 2 galaxy NGC\,1068 \citep{Antonucci85}. 
According to this model, 
the central engines of Seyfert 1 galaxies (hereafter Sy1s) can be seen directly, resulting in typical 
spectra with both narrow and broad emission lines produced in the kpc-scale narrow-line region (NLR) and the 
subparsec-scale broad-line region (BLR), respectively. On the other hand, the central engine and BLR of Seyfert 2 galaxies
(Sy2s) are obscured by the dusty toroidal structure that surrounds the BLR, preventing a direct view of the 
broad lines. Thus, the detection of these broad components in the polarized spectra of a Sy2 can be 
explained as scattered emission from the hidden BLR (HBLR).

After the unified model was proposed, a search for the HBLRs of Sy2s started, and indeed optical 
spectropolarimetry revealed broad components of H$\alpha$ in a significant fraction of them 
\citep{Miller90,Inglis93,Young96,Heisler97,Barth99,Barth99b,Alexander00,Moran00,Moran01,Lumsden01,Lumsden04,Tran01,Tran03}. 
However, up to date, $\sim$60--70\% of nearby Sy2s 
do not show a Sy1-type linearly polarized spectrum \citep{Moran01,Tran03}, and some authors proposed that these 
galaxies genuinely lack a Sy1 core, dubbed non-HBLR (NHBLR; \citealt{Tran01,Tran03,Gu02}). 
Another possibility extensively explored in the literature is that the non-detection of the broad components
is due to certain geometries/intrinsic properties of the scattering region, the torus and/or the host galaxy 
itself (e.g. \citealt{Miller90,Heisler97,Lumsden04,Marinucci12,Ichikawa15}). 

The majority of previous spectropolarimetric observations of Seyfert galaxies were obtained with 3-4 m class telescopes, 
except for those obtained with the 10 m Keck Telescopes \citep{Barth99,Barth99b,Moran01,Tran03}. Since then, tens of works reported 
differences between HBLRs and NHBLRs relying on old spectropolarimetry data. However, if we consider the high signal-to-noise 
ratios (S/N) required to detect the linear polarization signal typical of Seyfert galaxies ($\sim$1-5\%; \citealt{Antonucci85}), 
it is very likely that at least some NHBLRs could be misclassified due to insufficient data quality. Deeper spectropolarimetry 
data of complete samples of Seyfert galaxies obtained with 8--10 m class telescopes might hold the key to 
unveil the HBLR of nearby Sy2s. 

Here we present new linear optical spectropolarimetry of 15 Sy2s that reveal the presence of a HBLR 
in 11 galaxies at $\ge$4$\sigma$ (73\% of the sample).
Some of the targets were previously classified as NHBLRs, and for others there were no published 
spectropolarimetry data. Our results confirm that at least some galaxies currently treated as NHBLR in the literature
are misclassified, bringing previous publications reporting differences between HBLR and NHBLR objects into question.
Both detections and non-detections of HBLRs should always be reported in context with the achieved S/N.

\section{Sample, Observations and Data Reduction}
\label{observations}

Our sample consists of 15 Sy2s with log L$_{bol}\sim42.6-44.6$ erg~s$^{-1}$ for which we had constraints 
on the inclination angle of their tori (see Table \ref{tab1}). These constraints are 
either from previous spectral energy distribution (SED) fitting with torus models \citep{Alonso11,Ramos11,Ramos14},
from H$_2$O maser detections (disk-maser candidates only\footnote{Disk-maser detections require high nuclear obscuration, 
and therefore are generally associated with edge-on tori.}; \citealt{Zhang12}) and/or from X-ray observations of the Fe K$\alpha$ line 
(assuming that the accretion disk and the torus are coplanar; \citealt{Weaver98,Leighly99,Guainazzi10}). 
The sample includes 7 Compton-thick and 8 Compton-thin sources. 
We note that the Sy2 galaxies Centaurus A and NGC\,4945 were also included in the observed sample, but 
the S/N of their total flux spectra was insufficient to obtain a polarized spectrum. The combination of 
extreme nuclear obscuration\footnote{\citet{Marconi00a,Marconi00b} reported optical extinctions A$_V\sim14$ mag for the 
nuclei of Centaurus A and NGC\,4945.} and host galaxy dilution requires hours of infrared 
spectropolarimetric observations to obtain reliable polarized spectra for these galaxies (see e.g. \citealt{Alexander99}). 
Therefore, in the following we will refer to the 15 galaxies included in Table \ref{tab1}.
 
%\begin{deluxetable}{lcccccccccc}
%\tabletypesize{\scriptsize}
%\tablewidth{0pt}
%\tablecaption{Properties of the sample and detail of the observations.}
%\tablehead{
\begin{table*}
\centering
\footnotesize
\begin{tabular}{lccccccccccccc}
\hline
\hline
Galaxy & \multicolumn{3}{c}{Previous classification} 	& Axis ratio  & i$_{torus}$  &  Ref. & $\sigma_{torus}$ & Ref. & log n$_H$      & Compton  & log L$_{2-10}^{int}$ & log L$_{bol}$ & Ref. \\
       & Type   & Data &   Ref.                         & (b/a)       & (deg)        &       & (deg)            &      &(cm$^{-2}$)    &   thick	 & (erg~s$^{-1}$)  &	(erg~s$^{-1}$)       \\ 
\hline
Circinus                & HBLR  & $\surd$  	& a		& 0.44  &  90  &  m  & 60    & s1 &$>$24.5  & $\surd$   & 42.6 & 43.8 & 1       \\  
IC\,2560    		& \dots & $\times$  	& \dots  	& 0.63  &  90  &  n  & \dots &\dots&$>$24.5& $\surd$   & 41.8 & 43.1 & 2       \\  
IC\,5063    		& HBLR  & $\surd$   	& b,c  		& 0.68  &  80  &  o  & 60    & s2 &   23.4  & $\times$  & 42.8 & 44.0 & 1       \\  
NGC\,2110               & HBLR& $\surd$   	& d,e  		& 0.74  &  40$^{*}$&p& 45    & s3 &   22.5  & $\times$  & 42.5 & 43.9 & 3       \\  
NGC\,3081   		& HBLR  & $\surd$   	& f	 	& 0.78  &  71  &  q  & 75    & s4 &   23.9  & $\times$  & 42.5 & 43.6 & 1       \\  
NGC\,3281   		& NHBLR & $\times$   	& g		& 0.50  &  62  &  r  & 50    & s5 &   23.9  & $\times$  & 42.6 & 43.8 & 1       \\ 
NGC\,3393   		& NHBLR & $\times$      & h$^{\dag}$    & 0.91  &  90  &  n  & 67    & s6 &$>$24.5  & $\surd$   & 41.6 & 42.9 & 2,4     \\ 
NGC\,4388   		& HBLR  & $\surd$  	& i,j 		& 0.23  &  90  &  n  & 45    & s7 &   23.5  & $\times$  & 42.9 & 44.1 & 1       \\
NGC\,4941   		& NHBLR & $\times$   	& g       	& 0.54  &  76  &  s  & 50    & s8 &   23.8  & $\times$  & 41.3 & 42.6 & 5       \\  
NGC\,5135   		& NHBLR & $\surd$   	& k,l 		& 0.71  &  12  &  s  & 60    & s9 &$>$24.5  & $\surd$   & 43.1 & 44.4 & 1       \\  
NGC\,5506& NHBLR$^{\ddag}$ &$\surd$  & c    		        & 0.30  &  40  &  t  & 45    & s10 &   22.5  & $\times$  & 43.0 & 44.3 & 1       \\ 
NGC\,5643   		& NHBLR & $\times$  	& g  		& 0.87  &  74  &  q  & 60    & s11 &$>$24.5  & $\surd$   & 42.1 & 43.4 & 6,7     \\ 
NGC\,5728   		& NHBLR & $\surd$       & i$^{\S}$      & 0.57  &  90  &  n  & 60    & s12 &$>$24.5  & $\surd$   & 43.3 & 44.6 & 1       \\
NGC\,5793   		& \dots & $\times$ 	&\dots  	& 0.34  &  90  &  n  & \dots &\dots&$>$24.5  & $\surd$   & 42.1 & 43.4 & 8       \\ 
NGC\,6300   		& NHBLR & $\surd$ 	&  c   		& 0.66  &  77  &  u  & \dots &\dots&   23.3  & $\times$  & 41.8 & 43.1 & 9       \\
\hline
\end{tabular}
\caption{Columns 2, 3, and 4 indicate the previous classification of the galaxies as HBLR/NHBLR, the existence or not of previously published
spectropolarimetry data and corresponding references. Column 5 lists the axis ratio (b/a) of the galaxies from \citet{deVaucouleurs91}.
Columns 6--9 are the inclination angle of the torus (i$_{torus}$, with i$_{torus}$=90\degr~corresponding 
to edge-on orientations), torus width ($\sigma$), and corresponding 
references. Columns 10--14 indicate the hydrogen column densities measured from X-ray data, whether or not the source is Compton thick, 
the intrinsic 2-10 keV luminosities, the bolometric luminosities,
and corresponding references. 
$\dag$ HBLR according to \citet{Kay02}, but no data reported.
$\ddag$ Classified as obscured NLSy1 by \citet{Nagar02} and as HBLR by \citet{Tran01,Tran03}, but no spectrum reported.
$\S$ \citet{Young96} reported a rise in polarization around H$\alpha$, but did not detect a broad component.
* Based on [O III] and radio observations of NGC\,2110, \citet{Gonzalez02} and \citet{Rosario10} reported an orientation of 
the ionization cones and the radio jet $\sim$160-180\degr, which would be more consistent with an edge-on torus orientation. 
References. Spectropolarimetry: (a) \citet{Alexander00}; (b) \citet{Inglis93}; (c) \citet{Lumsden04}; (d) \citet{Moran07}; (e) \citet{Tran10}; 
(f) \citet{Moran00}; (g) \citet{Moran01}; (h) \citet{Wu11}; (i) \citet{Young96}; (j) \citet{Watanabe03}; (k) \citet{Heisler97}; 
(l) \citet{Lumsden01}. 
Torus inclination: (m) \citet{Greenhill03}; (n) \citet{Zhang12}; (o) \citet{Alonso11}; (p) \citet{Weaver98}; (q) \citet{Ramos14}; (r) \citet{Ramos11}; 
(s) this work; (t) \citet{Guainazzi10}; (u) \citet{Leighly99}.
Torus opening angle: (s1) \citet{Elmouttie98}; (s2) \citet{Schmitt03}; (s3) \citet{Rosario10}; (s4) \citet{Ferruit00}; (s5) \citet{Storchi92}; (s6) \citet{Cooke00}; 
(s7) \citet{Pogge88}; (s8) this work; (s9) \citet{Ichikawa15}; (s10) \citet{Wilson85}; (s11) \citet{Simpson97}; (s12) \citet{Wilson93}. 
X-ray data: (1) \citet{Marinucci12}; (2) \citet{Tilak08}; (3) \citet{Rivers14}; (4) \citet{Guainazzi05}; (5) \citet{Kawamuro16}; 
(6) \citet{Matt13}; (7) \citet{Annuar15}; (8) this work (see Appendix \ref{appendixB}); (9) \citet{Matsumoto04}.}
\label{tab1}
\end{table*}

We obtained linear spectropolarimetry with the FOcal Reducer and low dispersion Spectrograph 2 (FORS2) 
on the 8 m Very Large Telescope (VLT). The data were taken during the night of 2013 April 6th in visitor mode 
(Program ID: 091.B-0190A). The observing conditions were 
photometric and the seeing varied between 0.6\arcsec~and 1.6\arcsec. We used the 600B+22 grism and 
a slit width of 1\arcsec, which provide a dispersion of 0.75 \AA/pixel and a spectral resolution of R=$\lambda/ \Delta\lambda$=780. 
In order to obtain simultaneous observations of H$\beta$ and H$\alpha$, we shifted the position of the slit on the EV2 detector to 
cover the spectral range $\sim$4100--6800 \AA~with the 600B+22 grism. The pixel scale was 0.126\arcsec~pixel$^{-1}$.
%The pixel scale of the FORS2 EV2 detector is 0.126\arcsec. 

The exposure time in each half wave-plate position ranged between 200 and 300 depending on the target (see Table \ref{tab1}). 
The instrument position angle (PA) was set to parallactic, with the exception of the galaxies IC\,2560 and NGC\,2110, for which 
it was PA=0\degr. The data were reduced using routines written in IDL, following the method described by \citet{Martinez15}.
The reduction steps include bias subtraction, flat-fielding correction, wavelength calibration, sky subtraction, 
cosmic rays removal, and flux calibration. In the latter step, we used the spectrophotometric standard star LTT\,7379 and 
standard IRAF\footnote{IRAF is distributed by the National Optical Astronomy Observatory, which is operated by the Association 
of Universities for Research in Astronomy (AURA) under cooperative agreement with the National Science Foundation.} 
routines to derive and then apply the corresponding sensitivity function to the individual spectra. This 
G0 star was observed at the end of the night, using an exposure time of 50 s per half wave-plate position.

Since we are interested in the 
nuclear spectra of the galaxies, we extracted one dimensional spectra in an aperture coincident with the value of the seeing 
at the moment of each target observation (see Table \ref{tab1}). By doing this we also avoid contamination from the galaxy 
(i.e., dilution of the emission lines). Previous works 
generally employed larger apertures to minimize the amplitude of 
spurious features in the continuum (e.g. \citealt{Barth99,Moran01,Tran03}). However, since our aim is to detect the broad components 
of the recombination lines, we chose to reduce galaxy dilution by using nuclear apertures, which makes it easier to detect
the scattered lines. 
%By doing that we also minimize the contribution of dust within the host galaxies to the continuum polarization.

In order to calibrate the two beams produced by the FORS2 analyzer and to estimate 
the amount of instrumental polarization, we observed a linear polarization standard, the Hiltner 652 star, whose 
polarization degree and angle are known. According to the FORS2
webpage\footnote{http://www.eso.org/sci/facilities/paranal/instruments/fors/
inst/pola.html}, in the B-band the 
Hiltner 652 star has a degree of polarization of 5.72$\pm$0.02\% and a polarization angle of 179.8$\pm$0.1\degr. From 
our data, we measured 5.81$\pm$0.02\% and 180.1$\pm$0.1\degr~in the same filter, which we compute from the 
integrated Stokes parameters Q/I and U/I, weighted by the Johnson B filter transmission, as 
$P_B=[(Q/I)^2+(U/I)^2]^{1/2}$ and $\theta_B=0.5~arctan~(U/Q)$ (see \citealt{Martinez15} for further details). We define
$\theta$=0\degr~at the celestial N-S direction, increasing towards the East.
The agreement between the calculated and tabulated $\theta_B$ values implies that the instrument rotation has been chosen 
correctly in the data reduction, and the instrumental polarization 
in the B-band is $\sim$0.09\%. In order to check the latter, we also observed a zero polarization standard star, WD1620-391.
In the B- and V-bands we measured P$_B$=0.09$\pm$0.02\% and P$_V$=0.08$\pm$0.01\% respectively, confirming the value of the 
instrumental polarization.

Another potential contributor to the continuum polarization of the galaxies is Galactic dust \citep{Serkowski75}. 
In the case of our sample, the effect of interstellar polarization could be significant in the case of Circinus 
and NGC\,2110, because of their proximity to the Galactic plane. Using the \citet{Heiles00} catalog, 
\citet{Moran07} reported P$_V$=0.33\% and $\theta$=34.5\degr~for two stars at $\sim$12' from NGC\,2110. 
In the case of Circinus, \citet{Oliva98} measured P$_V\sim$1.8\% and $\theta\sim$68\degr~for a foreground star at 
4\arcsec~from the galaxy. We used those values to construct synthetic $Q$ and $U$ spectra
assuming a \citet{Serkowski75} curve, and subtracted them from the normalized Stokes parameters of NGC\,2110 and 
Circinus. The corrected polarization degree and position angle spectra are shown in Figures \ref{A1} and \ref{A4} of
Appendix \ref{appendixA}.

\begin{figure*}
\par{\includegraphics[width=8cm]{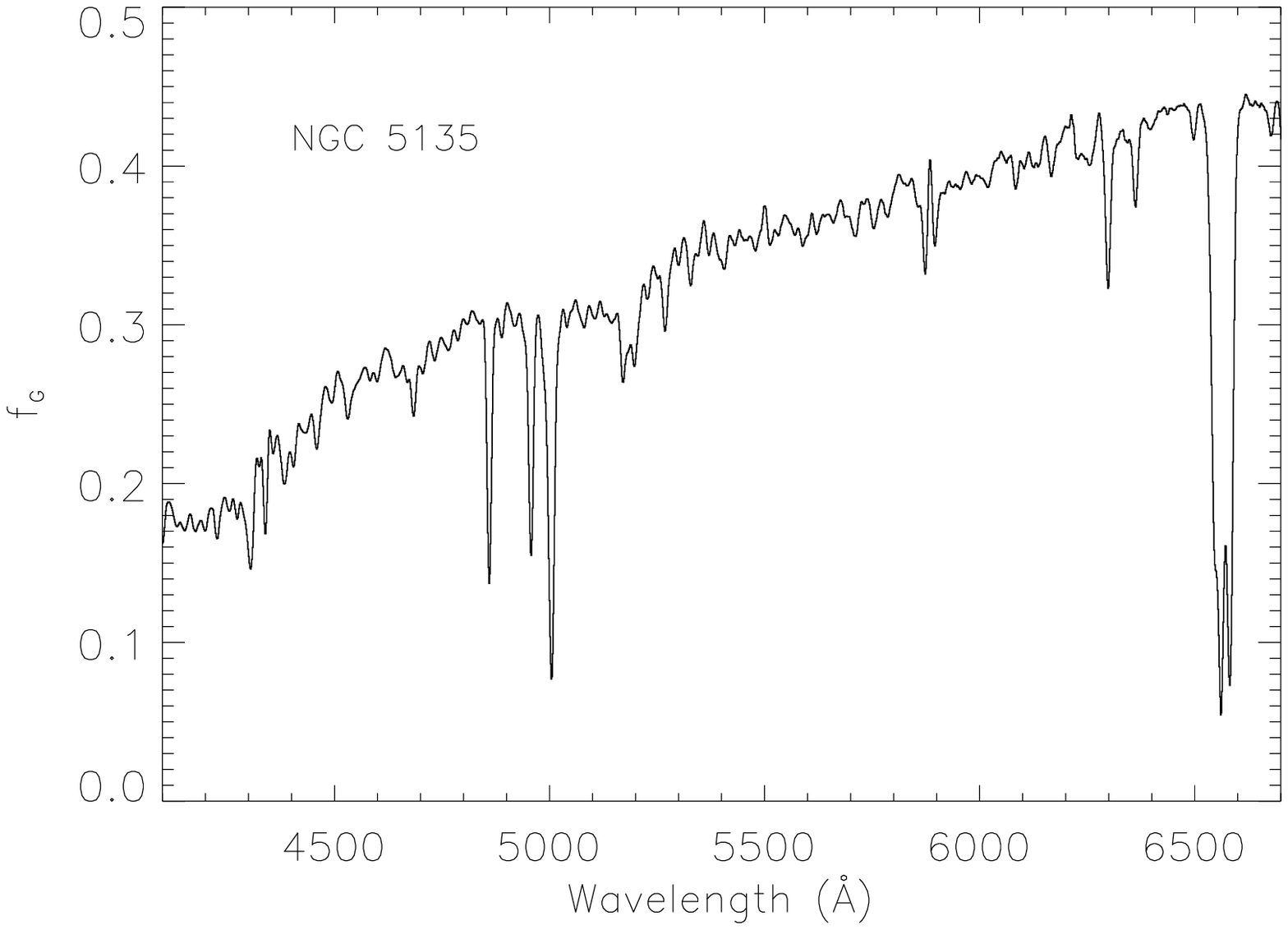}
\includegraphics[width=8cm]{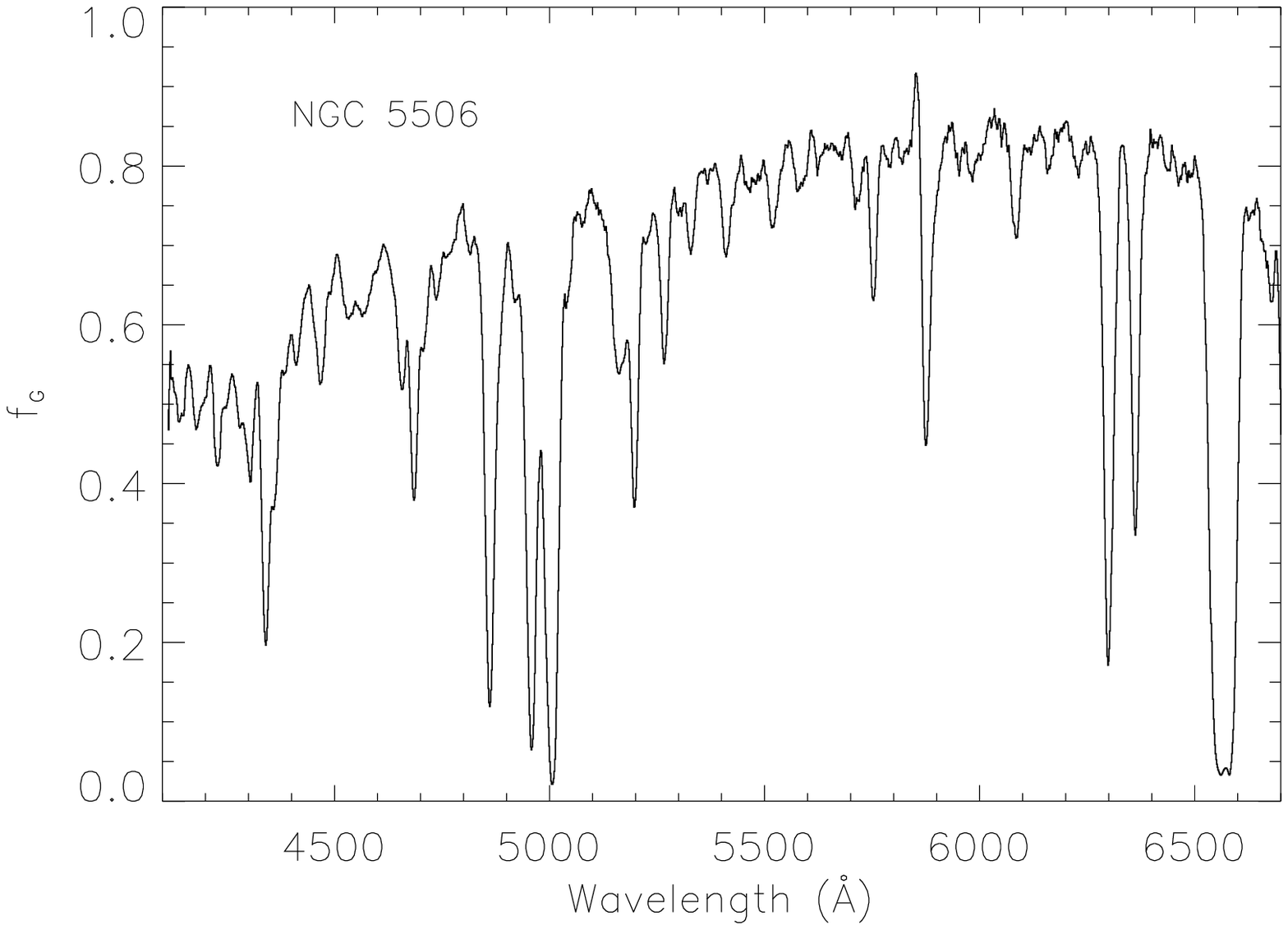}\par}
\caption{Galaxy fractions derived for NGC\,5135 and NGC\,5506. Details on the fits are given in Table \ref{tab2}.}  
\label{fig0}
\end{figure*}

Finally, we need to quantify the amount of diluting unpolarized starlight in the nuclear spectra of the galaxies.
To do that, we fitted the continua of our observed spectra with stellar population templates of different ages \citep{Vazdekis10}, 
including reddening, and a power-law ($F_\lambda \propto \lambda^{-\alpha}$) to account for the nonstellar nuclear continuum. 
We performed interactive fits until we successfully reproduced the stellar features of the underlying galaxy, and 
from the fitted templates we derived corresponding galaxy fraction spectra (f$_G(\lambda)$; 
see Figure \ref{fig0} for two examples). We used stellar templates of ages between 1 and 14 Gyr to reproduce the stellar 
features, and power-law indices ranging from 0 to 2. We note that the $\alpha$ values are in agreement with those derived from 
ultraviolet slope measurements of Sy2s \citep{Kinney91}.
Details of each individual fit, together with the galaxy fractions estimated at 4400 and 5500 
\AA, are reported in Table \ref{tab2}. f$_G(\lambda)$ decreases towards the blue as the starlight contribution becomes less important. 
We note that, although it is the best we can do, this method can produce extremelly different values of f$_G$ for the same objects 
(see \citealt{Tran95} and references 
therein). These differences do not only affect the derived polarization degrees, but also the wavelength dependence of the
polarization. 

As can be seen from Table \ref{tab2}, for the majority of the galaxies f$_G\sim$1 at 5500 \AA, and it ranges between
0.7 and 1 at 4400 \AA. The exceptions are NGC\,5135 and NGC\,5506 (see Figure \ref{fig0}). This implies that for all the 
galaxies but those two, the large amount of unpolarized starlight prevents to study the behaviour of the true polarization in the 
continuum.
Following \citet{Tran95}, the starligth dilution-corrected $Q_c$ and $U_c$ parameters are defined as:

\begin{equation}
\centering
Q_c=Q\frac{1}{1-f_G(\lambda)};   \hspace{1cm}   U_c=U\frac{1}{1-f_G(\lambda)}
\label{eq1}
\end{equation}
 
\begin{table}
\centering
\begin{tabular}{lccccc}
\hline
\hline
Galaxy & Age   & E(B-V) & $\alpha$ &  f$_G$ &  f$_G$	   \\
       & (Gyr) &        &  &  (4400 \AA) &  (5500 \AA)  \\
\hline
Circinus        & 1.0  & 1.274 & -0.1& 1.00 & 0.98       \\ 
IC\,2560        & 10.0 & 0.453 & 1.6 & 0.79 & 0.95       \\  
IC\,5063        & 5.0  & 0.344 & 3.6 & 0.66 & 0.96       \\  
NGC\,2110       & 14.1 & 0.646 & 1.1 & 0.84 & 0.97       \\  
NGC\,3081   	& 10.0 & 0.402 & 1.7 & 1.00 & 1.02       \\  
NGC\,3281   	& 14.1 & 0.506 & 1.5 & 0.98 & 0.99       \\ 
NGC\,3393   	& 14.1 & 0.272 & 1.7 & 1.00 & 1.00       \\ 
NGC\,4388   	& 11.2 & 0.579 & 2.1 & 0.93 & 1.00       \\
NGC\,4941   	& 12.6 & 0.572 & 2.0 & 0.91 & 1.00       \\  
NGC\,5135   	& 5.0  & 0.213 & 1.1 & 0.19 & 0.36       \\  
NGC\,5506       & 10.0 & 0.226 & 1.6 & 0.51 & 0.81       \\ 
NGC\,5643   	& 1.4  & 0.543 & 1.4 & 0.87 & 0.96       \\ 
NGC\,5728   	& 2.5  & 0.428 & 1.5 & 1.00 & 1.00       \\
NGC\,5793   	& 3.5  & 0.638 & 1.0 & 0.97 & 0.99       \\ 
NGC\,6300   	& 4.5  & 0.673 & 2.1 & 1.00 & 1.00       \\
\hline
\end{tabular}
\caption{Results of spectral decomposition. Columns 2, 3, and 4 correspond to the age of the stellar 
templates employed in the fits, reddening applied in each case (internal and interstellar), and power-law index (F$_{\lambda}\propto
\lambda^{-\alpha}$). Columns 5 and 6 list the
galaxy fractions derived from the fits at 4400 and 5500 \AA~respectively.}
\label{tab2}
\end{table}

Thus, we can only obtain realistic dilution-corrected Stokes parameters for NGC\,5135 and NGC\,5506 using Equation \ref{eq1}. 
We note that, while the polarization degree ($P$) changes substantially with this correction, the polarized flux 
($P \times F_{\lambda}$) is insensitive to it. The starligth dilution-corrected Stokes parameters and derived quantities 
are reported in Figures \ref{A9} and \ref{A10} of Appendix \ref{appendixA} and in Table \ref{tab3}.

\section{Results}

The sample studied here includes i) five Seyfert galaxies with previous detections of polarized broad components of the
H$\alpha$ line reported in the literature (also H$\beta$ in the case of NGC\,3081); ii) four galaxies classified 
as NHBLR based on previous spectropolarimetry data; and iii) six galaxies without any published spectropolarimetry data
(see Table \ref{tab1} for the references).
Four of the latter six galaxies (NGC\,3281, NGC\,3393, NGC\,4941, and NGC\,5643) are commonly treated as NHBLRs in the 
literature (e.g. \citealt{Zhang06,Wu11,Marinucci12,Koulouridis14}). In the case of NGC\,3281, NGC\,4941, and NGC\,5643, this 
classification comes from \citet{Moran01}, who claimed non-detection of polarized broad components in their spectra
from the Keck Telescopes, but never published the data. The origin of the classification of 
NGC\,3393 as a NHBLR is uncertain, and indeed \citet{Kay02} reported the detection of scattered broad H$\alpha$, 
but did not publish the data either. 
 
For each galaxy, we present the U/I and Q/I spectra, the degree of polarization and polarization angle, calculated as 
$P=[(Q/I)^2+(U/I)^2]^{1/2}$ and $\theta=0.5~arctan~(U/Q)$, and the flux-calibrated polarized and total spectra. 
The polarized spectra were calculated as the product of the total flux spectra and the degree of polarization (P$\times$F$_\lambda$). 
This is shown in Fig. \ref{fig1} for the galaxy NGC\,3393 and in Appendix \ref{appendixA} for the other 14 galaxies. 
Note that, as described in Section \ref{observations}, in the case of Circinus and NGC\,2110 we corrected the corresponding spectra
from the effect of interstellar polarization, and for NGC\,5135 and NGC\,5506, of unpolarized starlight dilution. 

\begin{figure*}
\centering
\includegraphics[width=15cm]{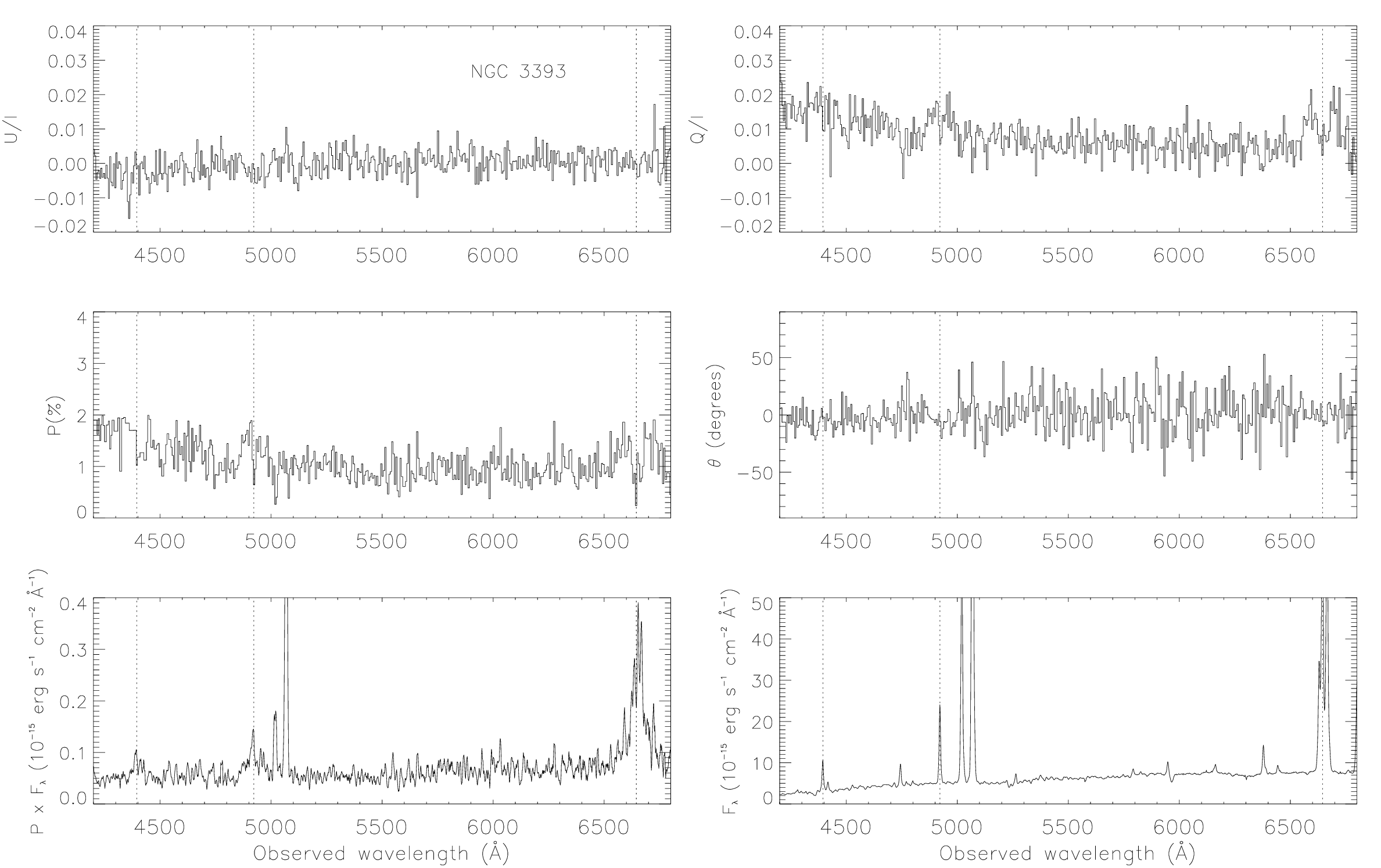}
\caption{Spectropolarimetry of the nucleus of NGC\,3393. Top panels correspond to U/I and Q/I, middle panels 
to the degree of linear polarization and polarization angle, and bottom panels to the 
observed flux-calibrated polarized and total spectra. The
polarized spectrum 
has been smoothed with a 5 pixels boxcar. Dotted vertical lines indicate the central position of H$\gamma$, 
H$\beta$, and H$\alpha$. Note the broad pedestals of these three recombination lines.}
\label{fig1}
\end{figure*}

\subsection{Significance of the detections}
\label{significance}

%The presence of scattered broad lines can be detected by comparing the emission line and continuum levels in the individual
%Stokes parameters spectra (top panels of Fig. \ref{fig1} and \ref{A1}--\ref{A13}); in the degree of polarization and polarization 
%angle (middle panels); and/or by comparing the line profiles in the total and polarized flux spectra (bottom panels).  

%Here we consider the detection of a scattered broad line confirmed when 1) it is detected either in the 
%Q/I or U/I 
%spectra {\bf at $\ge4\sigma$} over the continuum level, following the method described in \citet{Barth99};
%and 2) the residuals of the line fits not including the broad component show a significant excess when compared 
%with the residuals including it (see Figs.~\ref{fig2} and \ref{fig3}). {\bf We quantified the latter using the 
%Bayesian information criterion (BIC; \citealt{Schwarz78}).}

\begin{figure*}
\par{\includegraphics[width=8.5cm]{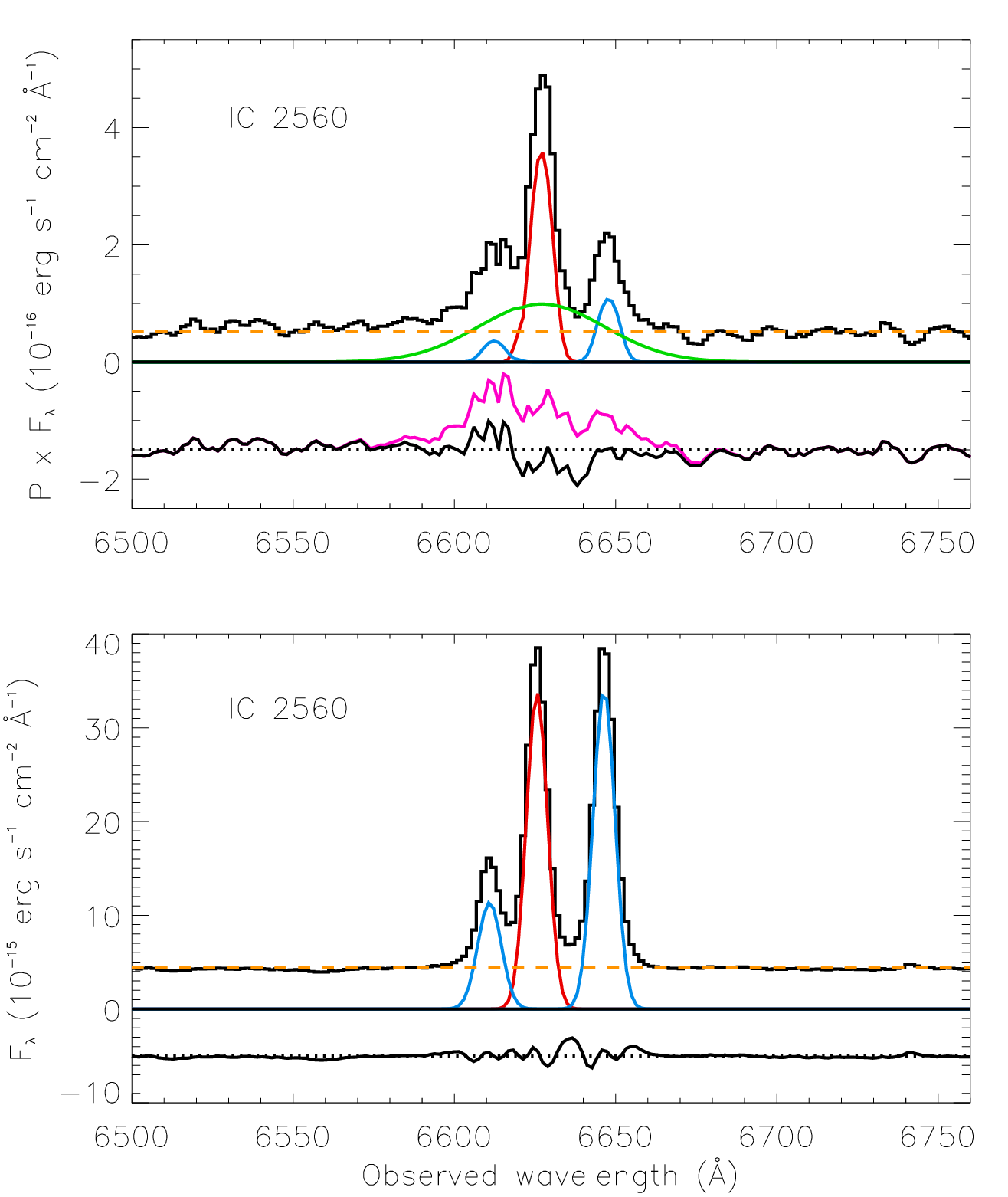}
\includegraphics[width=8.5cm]{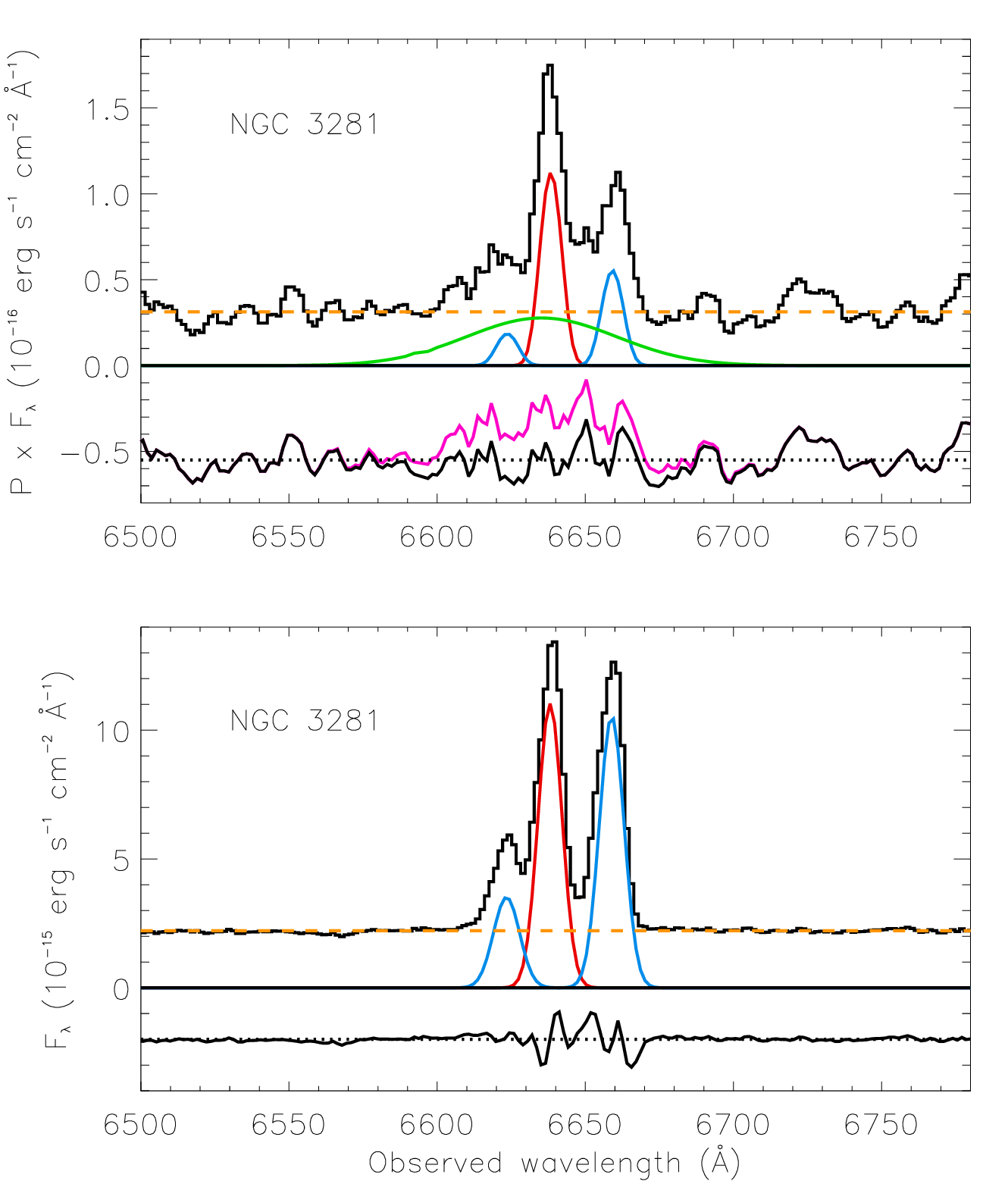}
\includegraphics[width=8.5cm]{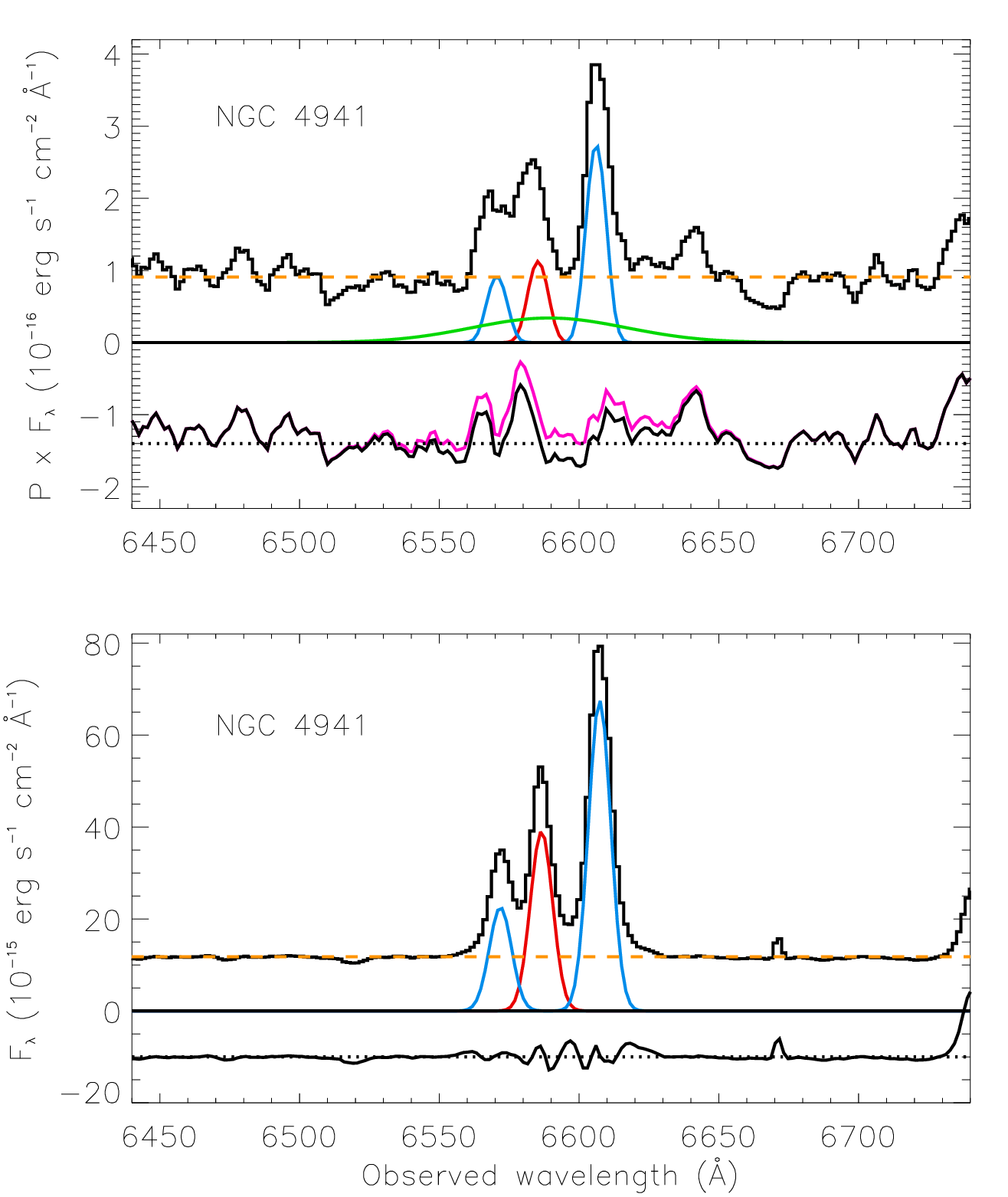}
\includegraphics[width=8.5cm]{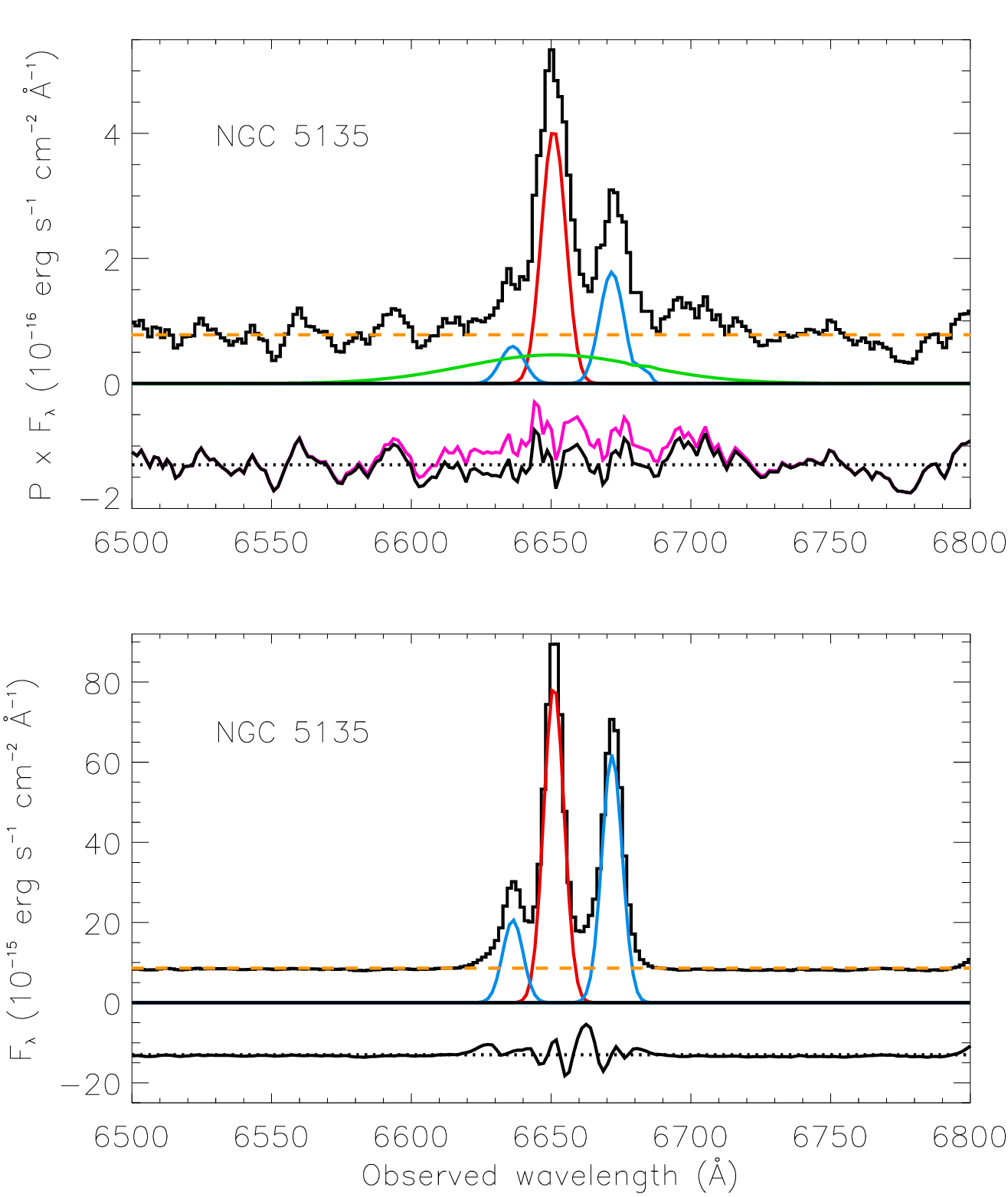}\par}
\caption{Comparison between the polarized and total flux H$\alpha$+[N II] profiles of the four galaxies 
in the sample with scattered broad components detected at $\le6\sigma$ and previously classified as 
NHBLRs or without previously published spectropolarimetry data.
Components of FWHMs$\sim$300--500 km~s$^{-1}$ have been fitted to reproduce the narrow emission lines, and 
broad components of FWHMs$\sim$2100--3400 km~s$^{-1}$ are necessary to reproduce the H$\alpha$ profiles in polarized flux. 
The insets at the bottom of each panel show the residuals of subtracting the fits from the spectra. In the case of 
the polarized spectra, we plot the residuals including and not including the broad component in the fit (black and pink
lines respectively).  
}\label{fig2}
\end{figure*}

\begin{figure*}
\par{\includegraphics[width=8.5cm]{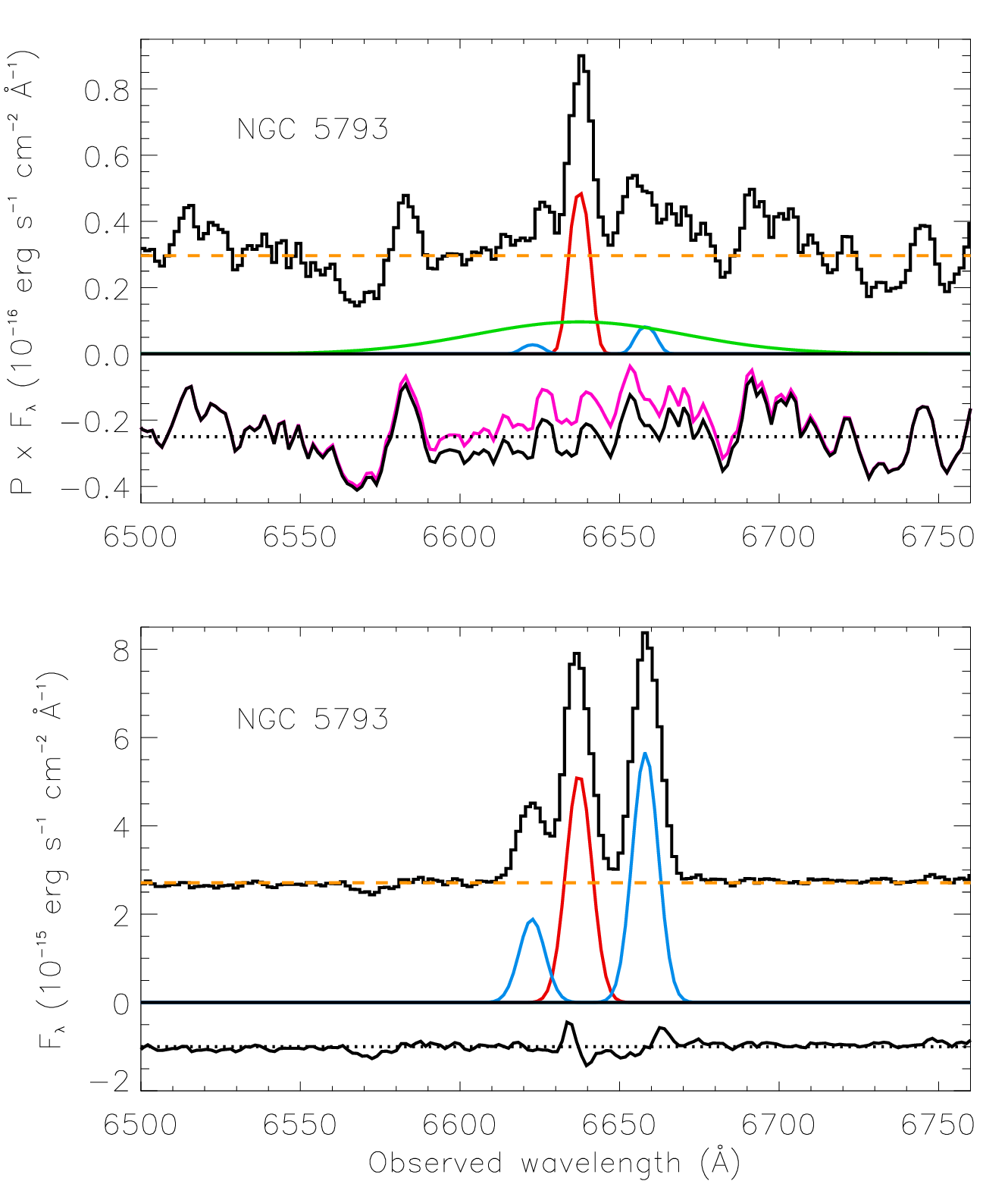}
\includegraphics[width=8.5cm]{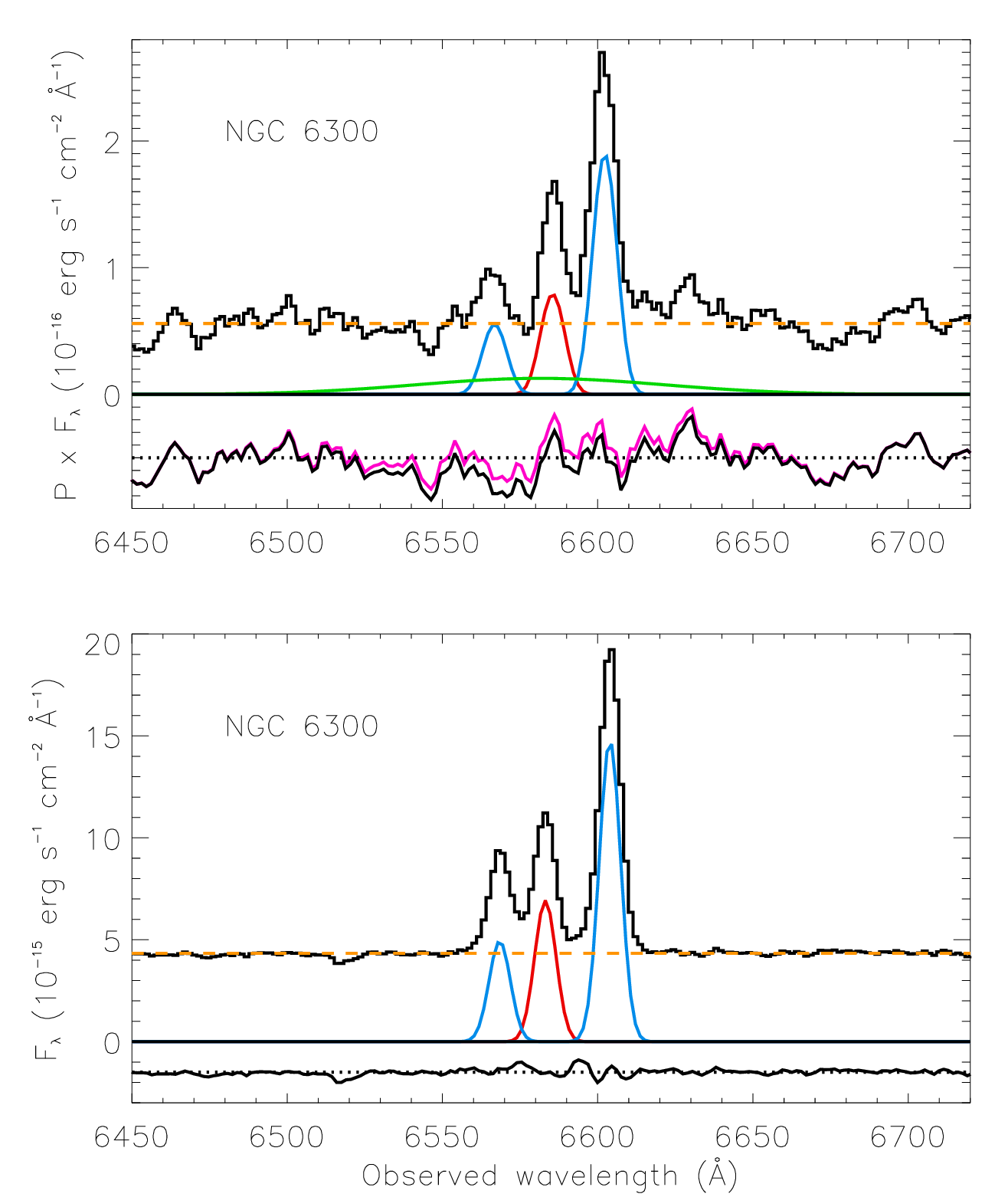}\par}
\caption{Same as in Fig.~\ref{fig2}, but for the galaxies NGC\,5793 and NGC\,6300. 
For NGC\,5793, we fitted a 
broad component of FWHM=3500$\pm$1200 km~s$^{-1}$ to the scattered H$\alpha$ profile, but the residual does not allow 
to confirm the presence of a HBLR in this galaxy (see Table \ref{tab3}).}  
\label{fig3}
\end{figure*}

%The polarized spectrum of each galaxy is the product of multiplying the degree of polarization 
%by the total-flux spectrum (P$\times$F$_\lambda$). 

We fitted the H$\alpha$, [N II]$\lambda\lambda$6548,6583 \AA, H$\beta$, and [O III]$\lambda\lambda$4959,5007 \AA~line 
profiles with Gaussians using the Starlink program {\sc dipso}. We used narrow components of 
full-width at half-maximum (FWHM) $\sim$300--500
km~s$^{-1}$, typical of NLR emission lines, to reproduce the line profiles in the total flux spectra. 
Normally, single Gaussian profiles are enough to accurately reproduce the narrow components, but in some cases 
two or three kinematic components were required (NGC\,2110, NGC\,3081, NGC\,4388, NGC\,5643, and 
NGC\,5728\footnote{In the case of NGC\,5728, three narrow components are necessary to reproduce the line profiles, as in
\citet{Son09}.}). Once we had a good fit to the 
total flux spectrum, we used the same input parameters (central wavelength of the lines and line width) 
and let them vary to fit the polarized flux spectrum (P$\times$F$_\lambda$). Finally, we checked whether we 
require broad components to reproduce the line profiles. 

We fitted broad H$\alpha$ and H$\beta$ components in polarized light 
for 10 of the galaxies, and just broad H$\alpha$ for NGC\,5793 and NGC\,6300. 
For the remaining three galaxies (NGC\,5506, NGC\,5643, and NGC\,5728) we do not find evidence for 
broad H$\alpha$ and H$\beta$ lines in the P$\times$F$_\lambda$ spectra presented here.

For the 12 galaxies in the sample for which we need broad components to reproduce the H$\alpha$ profiles detected in the polarized 
spectra, the FWHMs vary from 2100 to 9600 km~s$^{-1}$. 
In the case of H$\beta$, we fitted broad components of similar widths as those of H$\alpha$ 
for 9 galaxies and fixed the FWHM to match that of H$\alpha$ for NGC\,4941.
In the case of NGC\,5793 and NGC\,6300 we could not obtain reliable fits because of the low S/N of the H$\beta$ line.  
In Table \ref{tab3} we show the FWHMs of these broad components, corrected for instrumental broadening.

\begin{table*}
\centering
\footnotesize
\begin{tabular}{lccrrcccccccc}
\hline
\hline
Galaxy & Apt.  &  Exp. & $\Delta_U$ & $\Delta_Q$  & BIC &  P$_{H\alpha}$ & $\theta_{H\alpha}$ & P$_V$  & $\theta_V$ & H$\alpha$ FWHM & H$\beta$ FWHM & HBLR \\
       & (\arcsec) &  (s)     &            &             & H$\alpha$ &  (\%)          & (deg)             & (\%)   & (deg)      & (km~s$^{-1}$)   & (km~s$^{-1}$) & \\
\hline
Circinus$^{\dag}$       & 1.0   &  250	 &  4.4 & 4.9  & 228   & 1.77$\pm$0.06   &  19$\pm$2 & 1.71$\pm$0.02	& 17$\pm$3  & 2300$\pm$500  & 2800$\pm$2700	& $\surd$	\\  
IC\,2560    		& 1.3   &  300   & -4.7 & 1.1  & 363   & 0.82$\pm$0.08   & 104$\pm$2 & 0.77$\pm$0.03	& 91$\pm$1  & 2100$\pm$300  & 1800$\pm$1000	& $\surd$	\\  
IC\,5063    		& 1.6   &  200   &  4.0 & 0.3  & 3865  & 5.08$\pm$0.37   &  -1$\pm$3 & 3.54$\pm$0.23$^{\S}$ & -6$\pm$4  & 2800$\pm$200  & 2600$\pm$600& $\surd$	\\  
NGC\,2110$^{\dag}$      & 0.8   &  300   & 15.5 &-20.7 & 2214  & 2.54$\pm$0.07   &  75$\pm$2 & 0.67$\pm$0.03	& 87$\pm$4  & 9600$\pm$1400 & 9500$\pm$4800	& $\surd$	\\  
NGC\,3081   		& 1.0   &  200   & -2.4 & -5.9 & 59    & 0.54$\pm$0.08   &  89$\pm$2 & 0.29$\pm$0.03	& 86$\pm$8  & 2700$\pm$700  & 3000$\pm$2100	& $\surd$	\\  
NGC\,3281   		& 1.1   &  300   & -4.0 &  3.5 & 98    & 1.09$\pm$0.20   &  89$\pm$1 & 0.96$\pm$0.03	& 94$\pm$3  & 2700$\pm$700  & 3500$\pm$2200	& $\surd$	\\ 
NGC\,3393   		& 1.0   &  300   & -0.5 & 10.4 & 744   & 1.16$\pm$0.07   &   2$\pm$1 & 0.63$\pm$0.02	&  2$\pm$4  & 5000$\pm$600  & 5100$\pm$800	& $\surd$	\\ 
NGC\,4388   		& 0.9   &  200   & -7.2 &  1.1 & 157   & 0.78$\pm$0.11   & 128$\pm$19& 0.53$\pm$0.04	& 95$\pm$9  & 4500$\pm$1400 & 4300$\pm$1100	& $\surd$	\\
NGC\,4941   		& 0.8   &  300   &  4.5 & -0.9 & 10    & 0.24$\pm$0.06   & 38$\pm$11 & 0.06$\pm$0.02	& 18$\pm$7  & 2900$\pm$700  & 2900		& $\surd$	\\  
NGC\,5135$^{\ddag}$	& 0.8   &  300   & -4.0 & -0.9 & 75    & 0.67$\pm$0.08   & 105$\pm$7 & 0.66$\pm$0.03	&104$\pm$3  & 3400$\pm$1200 & 3700$\pm$1900	& $\surd$	\\  
NGC\,5506$^{\ddag}$     & 0.9   &  200   & -3.3 &  5.6 & \dots & 6.62$\pm$0.62   &  77$\pm$3 & 11.5$\pm$0.2 	& 75$\pm$2  & \dots         & \dots		& ?       	\\ 
NGC\,5643   		& 0.8   &  300   &  3.7 &  1.6 & \dots & 1.08$\pm$0.05   &  77$\pm$3 & 1.15$\pm$0.02  & 83$\pm$1  & \dots         & \dots             & $\times$	\\ 
NGC\,5728   		& 0.9   &  300   &  3.4 & -1.9 & \dots & 1.57$\pm$0.06   &  57$\pm$1 & 1.40$\pm$0.02  & 57$\pm$1  & \dots         & \dots 	        & $\times$	\\
NGC\,5793   		& 1.1   &  300   & -2.3 &  1.3 & 9     & 0.18$\pm$0.08   & 118$\pm$12& 0.08$\pm$0.03	& 86$\pm$13 & 3500$\pm$1200 & \dots		& ?		\\ 
NGC\,6300   		& 0.8   &  300   & -1.1 & -6.4 & 15    & 0.82$\pm$0.09   &  69$\pm$4 & 0.80$\pm$0.03	& 50$\pm$5  & 4200$\pm$2800 & \dots		& $\surd$	\\
\hline
\end{tabular}
\caption{Columns 2 and 3 list the apertures chosen for extracting the spectra and the exposure times in each of the 
four half wave-plate positions.
Columns 4 and 5 list the significances measured for H$\alpha$ in the Q/I and U/I spectra. Column 6 gives
the Bayesian information criterion (BIC) measured for the residuals from the H$\alpha$ fits including or not the broad component, with 
large BICs indicating stronger detections. 
Columns 7--12 are the degree of polarization and polarization angle in the H$\alpha$ bin and in the V-band, and the 
FWHM of the broad H$\alpha$ and H$\beta$ components, corrected from instrumental broadening. The FWHM of H$\beta$ 
has been fixed to match that of H$\alpha$ in the case of NGC\,4941. The last column indicates whether the presence of a HBLR has been confirmed from the results of this work. 
$\dag$ In the case of Circinus and NGC\,2110 we have corrected for interestellar polarization as described in Section \ref{observations}.
$\ddag$ The values here reported for NGC\,5135 and NGC\,5506 are corrected from unpolarized starlight, as described in Section \ref{observations}.
$\S$ P$_V$ was measured in the range 5800--6300 \AA~to avoid daylight polarization affecting blue wavelengths, as
IC\,5063 was observed at the end of the night.}
\label{tab3}
\end{table*}

To quantify the significance of the detections, we first measured the value of the
Stokes parameters by integrating in a bin width of 125 \AA~centred in H$\alpha$, following \citet{Barth99}. 
The 125 \AA~bin covers the rest-frame range 6500--6625 \AA, 
which is centered in H$\alpha$+[N II]. In the following 
we will refer to H$\alpha$ detections only, which are more and have higher S/N than those of H$\beta$.
In the case of the galaxies with very 
broad scattered lines (FWHM$>$100 \AA: NGC\,2110, NGC\,3393, and NGC\,4388), a broader H$\alpha$ bin of 180 \AA~was used 
(6480--6660 \AA). 
Those measurements were then compared with the corresponding Q/I and U/I continua, 
which we measured by integrating in two line-free adjacent bins of the same width, choosing 
the most adequate for each galaxy, as in \citet{Barth99}.  
In Table \ref{tab3} we report the significance level measured in the Q/I and U/I spectra for the individual galaxies ($\Delta_Q$ 
and $\Delta_U$), in standard deviations over the continuum level. 
As it can be seen from Figs.~\ref{fig1} and \ref{A1}--\ref{A14} in Appendix \ref{appendixA}, the Q/I and U/I spectra show 
spurious features at the 1--2$\sigma$ level, which correspond to polarization fringes \citep{Semel03}.
Therefore, here we only consider significant detections of scattered broad lines those having
$|\Delta_Q|\ge$4 and/or $|\Delta_U|\ge$4, which is the case of all the galaxies with broad lines detected but 
NGC\,5793 (see Table \ref{tab3}).  

Finally, to confirm the presence of a HBLR in the galaxies,
we compare the residuals from the fits to the polarized H$\alpha$ profiles including and not including the broad component. 
In Figs.~\ref{fig2} and \ref{fig3} we show the fits of the six galaxies 
in the sample with $|\Delta_U|$ and $|\Delta_Q|\le6\sigma$ previously classified as 
NHBLRs or without published spectropolarimetry data. To quantify the significance 
of these residuals, we have computed the Bayesian Information Criterion (BIC; \citealt{Schwarz78}) for all the galaxies, 
which is reported in Table \ref{tab3}. The BIC is used for model selection among a finite set of models, with the 
model with the largest BIC (in absolute value) resulting preferred. In our case, the BIC reported in Table
\ref{tab3} is the difference between the model fits with the broad component included minus the model without it. 
BICs larger than 10 can be considered very 
strong and those between 6 and 10 just strong \citep{Schwarz78}. The majority of the detections 
are very strong, although for NGC\,4941, NGC\,5793, and NGC\,6300 the BICs are considerably lower than for the rest 
of the galaxies.

%By putting together the significance of the detections and the BICs, 
%we confirm the presence of a HBLR in NGC\,4941 and NGC\,6300, although the BICs are not as significant as 
%for the rest of the HBLRs studied here. On the other hand, deeper observations are required
%to finally confirm NGC\,5793 either as hidden or non-hidden BLR.} 

%On the other hand, in the case of the galaxy NGC\,5728, the residuals shown in the top left panel 
%of Fig.~\ref{fig3} clearly reveal the presence of a broad H$\alpha$ component in the polarized flux spectrum. Thus, although 
%emission line polarization is detected at 3.4$\sigma$ over the continuum level\footnote{The significance of 
%the detection increases to 4.3$\sigma$ if a narrower H$\alpha$ bin of 70 \AA~is employed.}, we consider the presence 
%of a HBLR confirmed in this galaxy. 

Considering the previous criteria, here we report the detection of a HBLR for 11 of the 15
galaxies in our sample (73\%) at the 4$\sigma$ level.
We then confirm the existence of a HBLR for the five sources previously classified as such (Circinus, IC\,5063, 
NGC\,2110, NGC\,3081, and NGC\,4388) and unveil polarized broad components, for the first time, 
for another six galaxies. Provided that previous spectropolarimetry
of Seyfert galaxies revealed HBLRs in only $\sim$30--40\% of the observed targets \citep{Moran01,Tran03}, 
the detection of scattered H$\alpha$ emission in 73\% of the sample is noteworthy. 
Indeed, NGC\,5793 is probably a HBLR as well, but deeper spectropolarimetry data are required to confirm it. 
Finally, for the three galaxies whose fits do not require
scattered broad components (NGC\,5506, NGC\,5643, and NGC\,5728) we cannot discard the presence of 
a HBLR, but from the spectra presented here, broad components are not necessary to reproduce the 
line profiles in polarized flux. Perhaps the presence of two or more kinematic components in these galaxies
prevents the detection of faint and broad lines \citep{Nagar02,Son09,Cresci15}.
Higher S/N observations than those presented here of a complete
sample of Sy2 galaxies are required to either confirm or discard the ubiquitous existence of a HBLR in 
this type of AGN.

\subsection{Polarization degree, polarization angle and FWHM of the broad components}

In Table \ref{tab3} we also report the values of the polarization degree and the polarization angle measured in the same 
H$\alpha$ bins employed for calculating $\Delta_Q$ and $\Delta_U$ (P$_{H\alpha}$, $\theta_{H\alpha}$) and in the Johnson
V-band filter (P$_{V}$, $\theta_{V}$). Uncertainties were calculated as described in \citet{Bagnulo09} 
and \citet{Martinez15},
and they include photon noise and background subtraction errors. 
We note that in the case of galaxies with low continuum polarizations, 
the values of the polarization degree reported in Table \ref{tab3} are significantly lower than those that can be 
visually inferred from Figs. \ref{fig1} and \ref{A1}--\ref{A14}. This is because P$_{H\alpha}$ and P$_{V}$ are calculated
from Q/I and U/I integrated in the previously mentioned filters, which results in lower values than those that 
we would obtain by, for example, fitting a straight line to the continuum. 

The majority of P$_{V}$ values reported in Table \ref{tab3} are well above the instrumental polarization 
that we measured in this band 
(0.08$\pm$0.01\%), except those of NGC\,4941 and NGC\,5793, which can be explained by instrumental polarization. 
%These are the two objects for which we cannot confirm the presence of a HBLR based on the data presented here. 
In general, we detect significant enhancements in the polarization degree near H$\alpha$ with respect to the 
continuum, confirming the presence of scattered H$\alpha$ emission \citep{Lumsden01,Moran00,Tran10}. 
On the other hand, some galaxies show P$_{H\alpha}$ values which are consistent with P$_{V}$ considering the errors. 
The
exception is NGC\,5506, for which we measure P$_{H\alpha}$=6.6$\pm$0.6\% and P$_{V}$=11.5$\pm$0.2\%. Thus, for this galaxy we observe a 
depolarization around the position of H$\alpha$, as also reported by \citet{Lumsden04}.

Middle panels of Figs. \ref{fig1} and \ref{A1}-\ref{A14} show that the continuum polarization 
is roughly constant with wavelength for the majority of the galaxies, although there is a rise below $\sim$4500 \AA, where 
dilution by starlight decreases (see Section \ref{observations}). A constant polarization degree would 
be associated with electron scattering. However, as shown in Section \ref{observations}, the large amount of unpolarized starlight dilution 
that affects the majority of the galaxies studied here (see Table \ref{tab2}), as well as the uncertainties inherent to 
the method employed for correcting it, prevent a detailed 
analysis of the polarization mechanisms at play.

\begin{figure*}
\par{\includegraphics[width=6cm,angle=90]{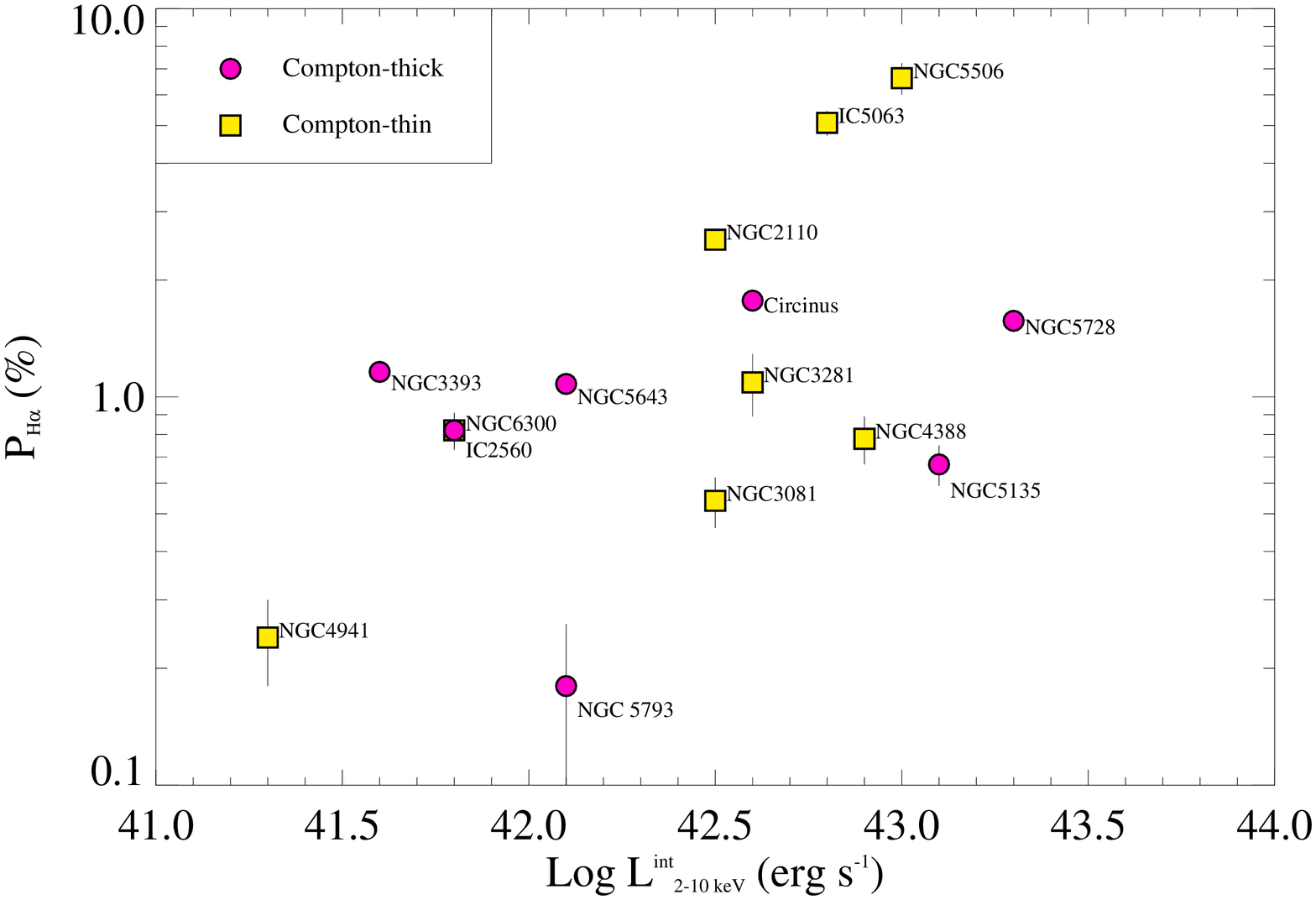}
\includegraphics[width=6cm,angle=90]{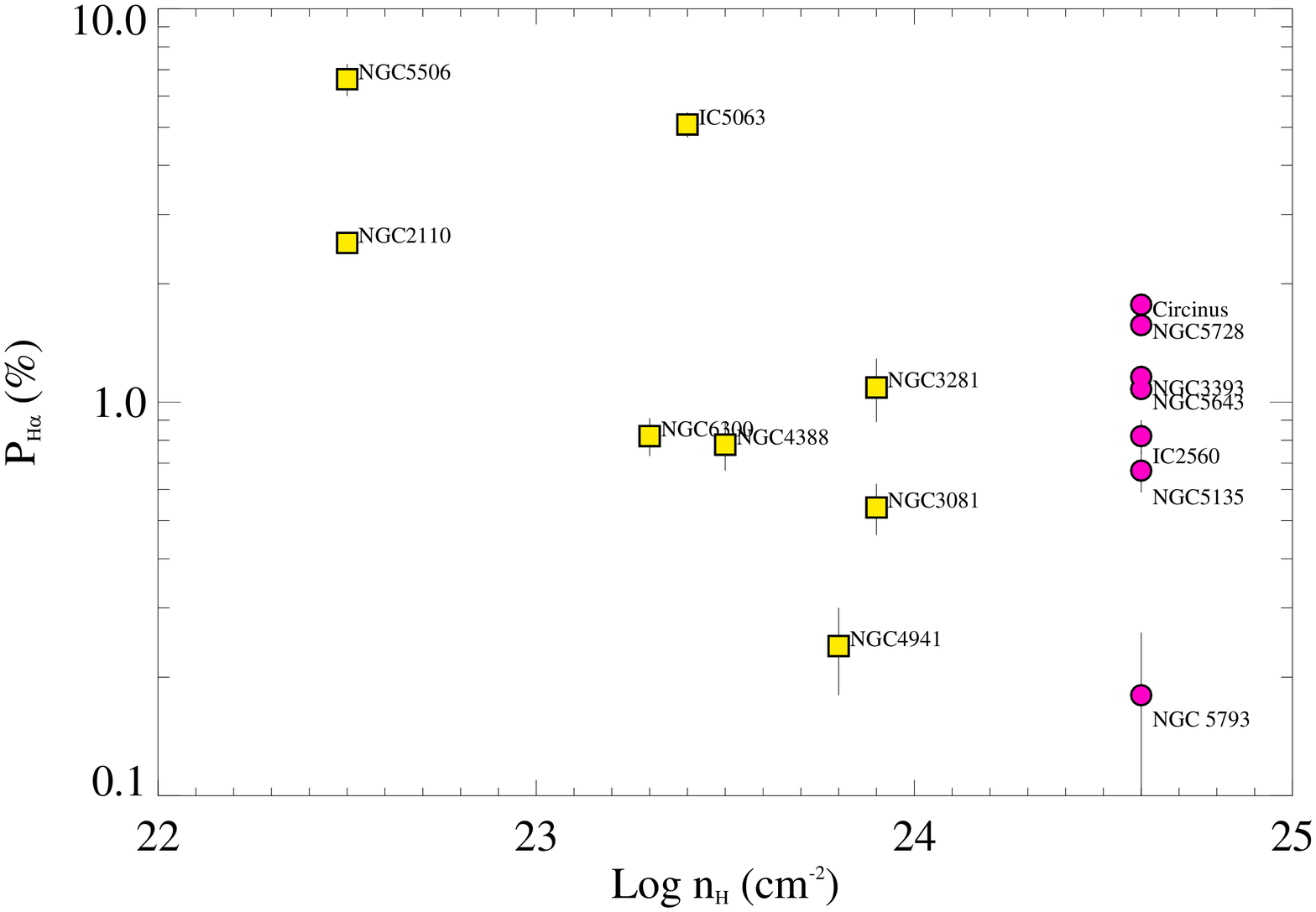}
\includegraphics[width=6cm,angle=90]{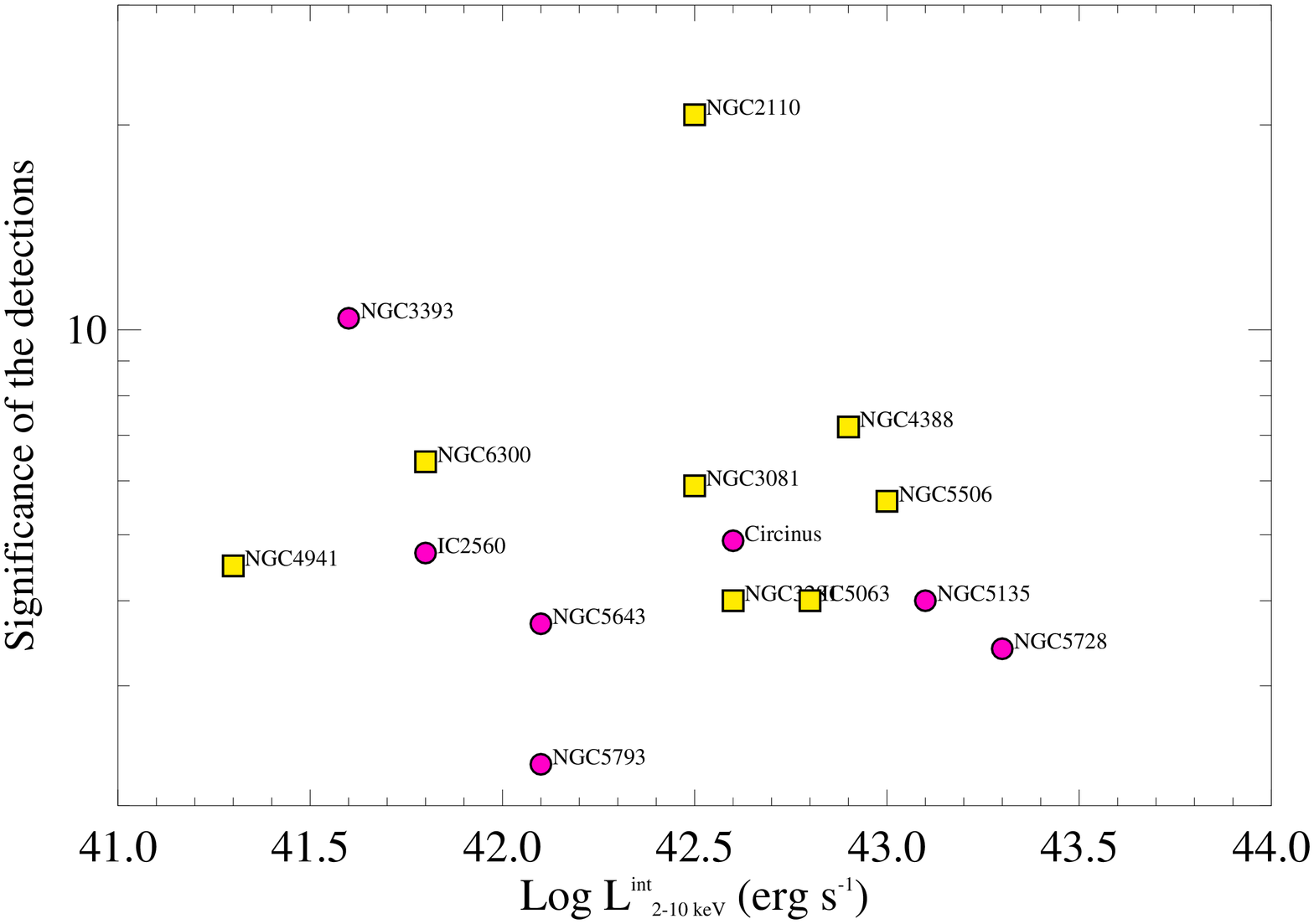}
\includegraphics[width=6cm,angle=90]{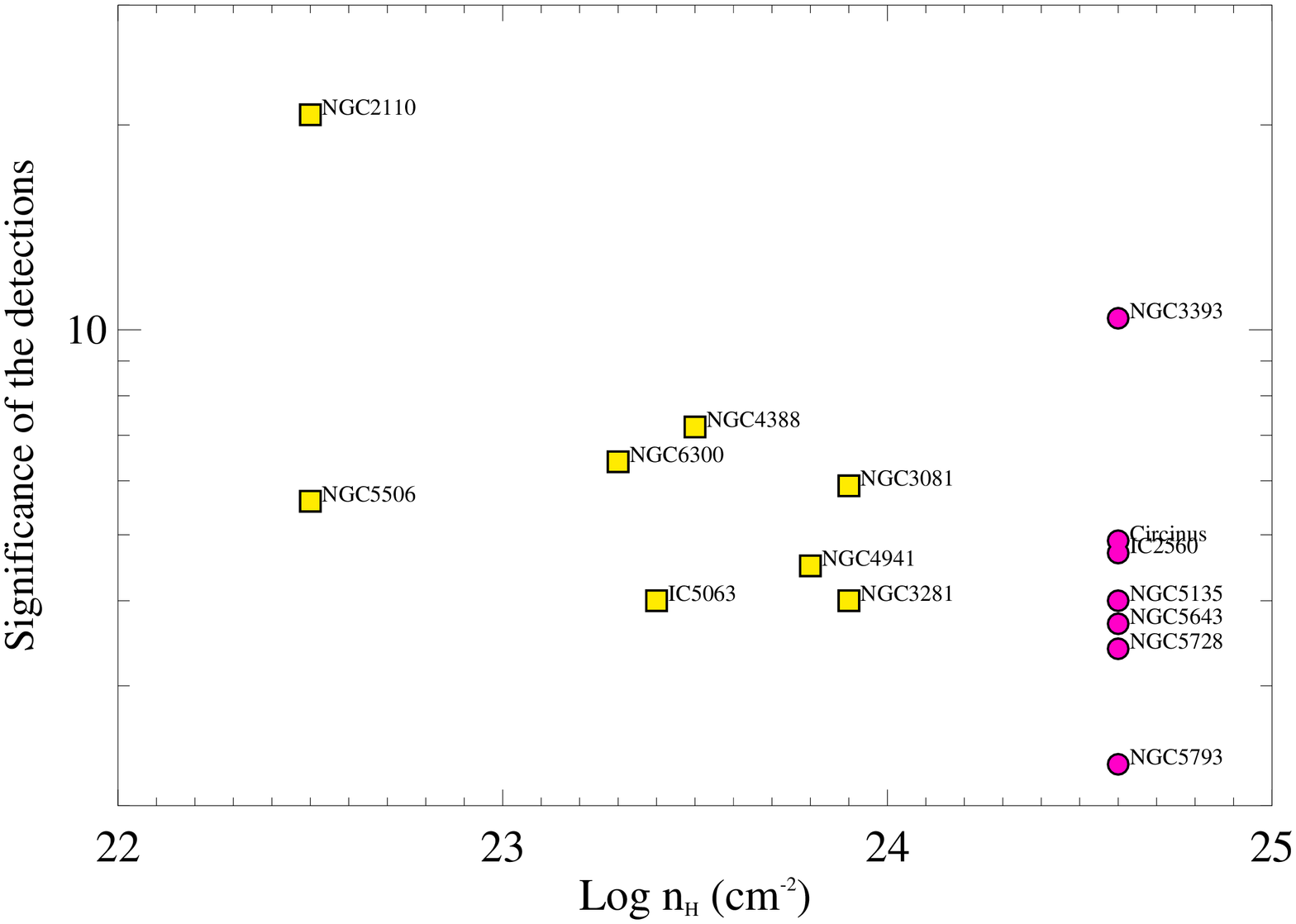}\par}
\caption{Left panels: polarization degree measured in the H$\alpha$ bin (P$_{H\alpha}$) and significance of the scattered H$\alpha$ detections ($|\Delta_Q|$ or
$|\Delta_U|$) versus intrinsic 2--10 keV luminosity. 
Right panels: same as in the left panels but versus the hydrogen column density measured from X-ray data. Pink 
circles and yellow squares are Compton-thick and Compton-thin sources respectively.}
\label{fig4}
\end{figure*}

In the top left panel of Figure \ref{fig4} we show the polarization degree measured in the H$\alpha$ 
bin versus the intrinsic 2--10 keV luminosities reported in Table \ref{tab3}. In spite of the reduced size of the 
sample considered here and the uncertainties associated with the polarization degree described in Section \ref{observations}, 
we find that relatively large P$_{H\alpha}$ values are measured for galaxies with log L$_{2-10}^{int}\ge$42.5 erg~s$^{-1}$.
However, if we look at the significance of the scattered H$\alpha$ detections ($|\Delta_Q|$ or $|\Delta_U|$ from Table \ref{tab3}), 
the previous trend disappears (see bottom left panel of Figure \ref{fig4})\footnote{The detection significances 
plotted in Figure \ref{fig4} are equivalent to the
ratio between the polarization degrees measured in the H$\alpha$ and V-band filters (P$_{H\alpha}/P_V$).}. 
We find something similar if we look at the hydrogen column density instead (see right panels of Figure \ref{fig4}). While higher 
polarization degrees are measured in galaxies with lower n$_H$, there is no correlation when we look at the significance of the 
detections. 

%%The case of NGC\,5793 is remarkable: from our analysis of {\emph Chandra} and {\emph XMM-Newton} data available for this source,
%which we describe in Appendix \ref{appendixB},
%we conclude that this galaxy is a Compton-thick low-luminosity Seyfert (log L$_{bol}$=41.5 erg~s$^{-1}$). This bolometric
%luminosity is indeed below the limit predicted by \citet{Elitzur09} for the disappearance of the BLR and the torus, further 
%supported by \citet{Gonzalez15} using infrared data from {\emph Spitzer}. Deeper spectropolarimetry of NGC\,5793 is then
%required to then confirm the detection of a hidden BLR in this galaxy. }

%%A similar trend can be seen from the right panel of Figure \ref{fig4}, where we represent the nuclear
%12 $\micron$ fluxes of the galaxies from \citet{Asmus14}\footnote{The 12 $\micron$ fluxes of IC\,2560 and NGC\,5793 correspond
%to IRAS data, as they are not included in \citet{Asmus14}.}. In general, the larger the mid-infrared nuclear fluxes 
%the higher the polarization degree. We find the same results if we consider P$_V$ instead of P$_{H\alpha}$.}

Regarding the polarization angles, we observe rotations across the Balmer lines 
in some of the galaxies. For example, the middle right panels of Figs. \ref{A3}, \ref{A4}, \ref{A7}, and \ref{A10} clearly show 
different $\theta$ features at the position of H$\alpha$ and, to a lesser extent, H$\beta$,
for the galaxies IC\,5063, NGC\,2110, NGC\,4388, and NGC\,5506. Moreover, if we look at the $\theta_V$ and $\theta_{H\alpha}$ values 
reported in Table \ref{tab3}, rotations across H$\alpha$ become evident for many of the of the targets.

%and the polarization degree measured in a narrow spectral region centred in H$\alpha$ (P$_{H\alpha}$; $\Delta\lambda$=110 \AA) and 
%in the V-band continuum (P$_V$; $\Delta\lambda$=450 \AA). As expected, we detect 
%an enhancement in the polarization degree near H$\alpha$ with respect to the 
%continuum for all the galaxies \citep{Lumsden01,Moran00,Tran10}. The only exception is NGC\,5135, 
%although P$_{H\alpha}\ga P_V$ is still compatible with the errors. 

As explained in Section \ref{significance}, we fitted broad components to the H$\alpha$ and H$\beta$ profiles
detected in the P$\times$F$_{\lambda}$ spectra of 10 galaxies, and just broad H$\alpha$ for another two. 
The broadest components that we measured are 
those of NGC\,2110, NGC\,3393, NGC\,4388, and NGC\,6300 (FWHMs $\sim$4200--9600 km~s$^{-1}$). These broad
lines were already reported by \citet{Moran07} and \citet{Tran10} for NGC\,2110 and by \citet{Young96} for 
NGC\,4388. A polarized spectrum of NGC\,6300 was presented by \citet{Lumsden04}, but they did not find any 
evidence for broad H$\alpha$, although they could not rule out the presence of a HBLR in this galaxy 
because of the low S/N of the data. For NGC\,3393, here we report the first polarized spectrum of the 
galaxy, which reveals clear broad H$\alpha$ and H$\beta$ components of $\sim$5000 km~s$^{-1}$. This broad
component is also detected in H$\gamma$, but the emission line is too faint to obtain a reliable fit (see
Fig. \ref{fig1}).
For the rest of the galaxies with broad components detected, the FWHMs range between 1800 and 3700 km~s$^{-1}$.

For NGC\,5506 we fitted intermediate-width H$\alpha$ and H$\beta$ components of FWHMs$\sim$1700
km~s$^{-1}$ to reproduce the line profiles detected in both polarized and total flux spectra. 
\citet{Nagar02} reported the detection of a broad
pedestal of 1800 km~s$^{-1}$ in the Pa$\beta$ line, using near-infrared spectroscopy, and classified the galaxy as an 
obscured NLSy1. The Stokes parameters, polarization degree, and polarization angle of NGC\,5506, which 
are corrected from starlight dilution, show features at the position of H$\alpha$ and H$\beta$ (see Figure 
\ref{A10} and Table \ref{tab3}), but we do not need a broader component to reproduce the line profiles in polarized flux.
Indeed, the $\Delta$Q that we measure for NGC\,5506 is 5.6 (see Table \ref{tab3}), indicating a significant detection.
This galaxy has a low column density as measured from X-ray data, but it is nearly edge-on (see Table \ref{tab1}) 
and it has a prominent dust lane, which might complicate the detection of the broad lines. 

Finally, two and three narrow components were needed to reproduce the line profiles of NGC\,5643 \citep{Cresci15} 
and NGC\,5728 \citep{Son09} respectively. Again, no broad components are required to fit the permitted lines in 
polarized flux, and furthermore, the detections of scattered H$\alpha$ emission in the 
Stokes spectra are not significant (see Table \ref{tab3}).

%\begin{figure}[!h]
%\centering
%\epsscale{1.04}
%\plotone{f4.eps}
%\caption{Same as in Fig. \ref{fig2}, but for the galaxy NGC\,5506. 
%Components of FWHM=400$\pm$10 km~s$^{-1}$ have been fitted to reproduce the narrow emission lines, and 
%a broad component of FWHM=1750$\pm$50 km~s$^{-1}$ is necessary to reproduce the H$\alpha$ profile in both 
%polarized and total flux. 
%}\label{fig4}}
%\end{figure}

\section{Discussion}
\label{discussion}

Based on the new VLT/FORS2 spectropolarimetric observations presented here,
we report the detection of a HBLR for five Sy2 galaxies previously classified as NHBLRs and one 
galaxy without any previous classification (IC\,2560; see Tables \ref{tab1} and \ref{tab3}). 
This result stresses the need for revisiting the current classification of NHBLRs. 
The VLT/FORS2 spectra studied here are not particularly deep, as the integration times employed 
were relatively short (see Section \ref{observations}), and yet, they allowed us to detect
scattered broad H$\alpha$ lines at $\ge4\sigma$ 
for 11/15 galaxies in the sample (73\%). Moreover, we report a tentative detection of a HBLR in 
the galaxy NGC\,5793, although deeper observations are needed to confirm it.

%Even if we only consider a more restrictive detection level of 5$\sigma$ for the scattered broad lines, we 
%reclassify four NHBLRs as HBLRs (NGC\,3393, NGC\,5506, NGC\,5643, and NGC\,6300).}

The new data presented here might put published results based on previous HBLR/NHBLR classifications into 
question, as we have proved that at least some NHBLRs are misclassified. Detections and non-detections 
of HBLRs should always
be put in context with the achieved S/N of the observations. Deep and high quality spectropolarimetry data 
of complete samples of Sy2 galaxies are required before continue searching for trends that might 
explain the non-detection of polarized broad lines.

We detect polarized broad components in both Compton-thick and Compton-thin objects (see Table \ref{tab1}). 
Several authors claimed that the detection of HBLRs might depend on 
the column density of the obscuring material (n$_H$), with a higher percentage of HBLRs found in Compton-thin 
sources \citep{Gu01,Lumsden04,Shu07}. However, for the galaxies in our sample we do not find a correlation between 
n$_H$ and the detectability of the polarized broad lines (see bottom right panel of Figure \ref{fig4}). Indeed, 
in the case of the three galaxies for which we do not detect broad components 
two are Compton-thick (NGC\,5643 and NGC\,5728) and NGC\,5506 has a low column density (see Table \ref{tab1}).
According to the top right panel of Figure \ref{fig4}, however, we find that the largest polarization degrees are 
measured for the galaxies with lower n$_H$. 
We find a similar behavior if we look at the intrinsic 2--10 keV luminosities (left panels of Figure \ref{fig4}). 
Relatively high polarization degrees are measured for galaxies with log L$_X^{int}\ge 42.5~erg~s^{-1}$, and, in general, the more luminous the 
galaxies, the higher the polarization degrees. However, this trend vanishes when we plot the significance of the detections instead (see 
bottom left panel of Figure \ref{fig4}). High bolometric luminosities and low column densities are associated with higher
polarization degrees, but that does not necessarily imply more HBLR detections.

On the other hand, the so-called ``true Sy2s'' are low-column density galaxies supposed to intrinsically lack a BLR
\citep{Tran11}. The majority of these sources are low-luminosity AGN (e.g. \citealt{Tran11,Stern12a}), for which 
the absence of scattered broad lines can be explained e.g. by the disappearance of the BLR and the torus predicted in the 
disc-wind scenario \citep{Elitzur09}, or just by dilution due to the stellar emission in the host galaxy. However, more recently, at least 
two high Eddington-ratio AGN of bolometric luminosities
similar to those considered here have been suggested as ``true Sy2s'' candidates based on X-ray and optical 
observations \citep{Ho12,Miniutti13}, but spectropolarimetry data are needed to confirm the intrinsic lack of BLR in these
galaxies.
 
%In Figure \ref{fig4} we show that the polarization degrees measured in the H$\alpha$ bin appear to increase with 
%the intrinsic X-ray luminosity and with the mid-infrared nuclear flux. We observe the same effect if we look at P$_V$
%instead.   
There is no correlation between the detection of scattered broad lines and galaxy inclination (b/a) for the galaxies in 
our sample.
Among the four galaxies which are nearly edge-on in the sample discussed here (Circinus, NGC\,4388, NGC\,5506 and NGC\,5793), 
we have confirmed and tentative HBLR detections.

Finally, we do not find the correlation we were expecting when we selected the sample 
between the detectability of the polarized broad lines and the 
torus properties. As explained in Section \ref{observations}, we have torus inclination constraints for all the 
galaxies, but we do not find any correlation with the properties of the polarized spectra 
(P, $\theta$ and FWHM of the broad lines).
%For example, according to the unified model, the scattering of the 
%nuclear continuum should occur along a direction parallel to the axis of the obscuring torus \citep{Moran07}. Following this
%scheme, the torus inclination angles reported in Table \ref{tab1} should be similar to the polarization angles given in 
%Table \ref{tab3}. We find that this is the case for only six of the galaxies in our sample (e.g. IC\,2560, NGC\,3081, NGC\,4388,
%NGC\,5643, NGC\,5793, NGC\,6300). We can also add the NGC\,2110 to the previous list 
%if we consider the more likely edge-on torus orientation that is consistent 
%with the axis of the radio jet and the ionization cones; see Table \ref{tab1}).}
A possible explanation for this lack of correlation is that the scale height of the 
scattering region (either dust or electrons) is large enough for the scattered light to be detected independently of 
the torus inclination (see \citealt{Ichikawa15} for further discussion). 
Furthermore, we note that some of the torus inclination constraints that we have come from SED fitting with 
torus models \citep{Alonso11,Ramos11,Ramos14} and therefore are model-dependent. Besides, we are biased to 
large inclination angles (12 galaxies have torus orientations $>60\degr$, i.e. close to edge-on, and the other three
$<45\degr$), making it difficult to distinguish any trend. 

We did not find any correlation with the 
torus width $\sigma$ either, which we derived from the opening angle of the ionization cones for the majority of the 
galaxies and from SED modelling with clumpy torus models for NGC\,4941 and NGC\,5135 (see Table \ref{tab1}). 
Among the possible explanations for the lack of a HBLR in some Sy2s, \citet{Tran03} proposed that the obscuring medium in 
these objects is not a torus but something more extended. However, as noted by the latter author, this is difficult to 
reconcile with the presence of ionization cones in galaxies such as NGC\,5728 and NGC\,5643 among others. Another possibility
are the different torus widths or covering factors (see e.g. \citealt{Ramos11}), which can make easier or more difficult the 
detection of HBLRs. For the 12 galaxies in 
our sample with $\sigma$ measurements available, which range between 45\degr~and 75\degr, we do not find any trend with 
the polarization properties presented here. Spectropolarimetric observations of a larger sample of Sy2s
are required to further investigate the influence of torus properties on the detection of the scattered broad lines.

\appendix
\section{Spectropolarimetry of the individual galaxies}
\label{appendixA}

Here we include the spectropolarimetry data of all the galaxies in the sample discussed here, with the exception of 
NGC\,3393, for which the data is shown in Fig. \ref{fig1}.

\begin{figure*}
\centering
\includegraphics[width=15cm]{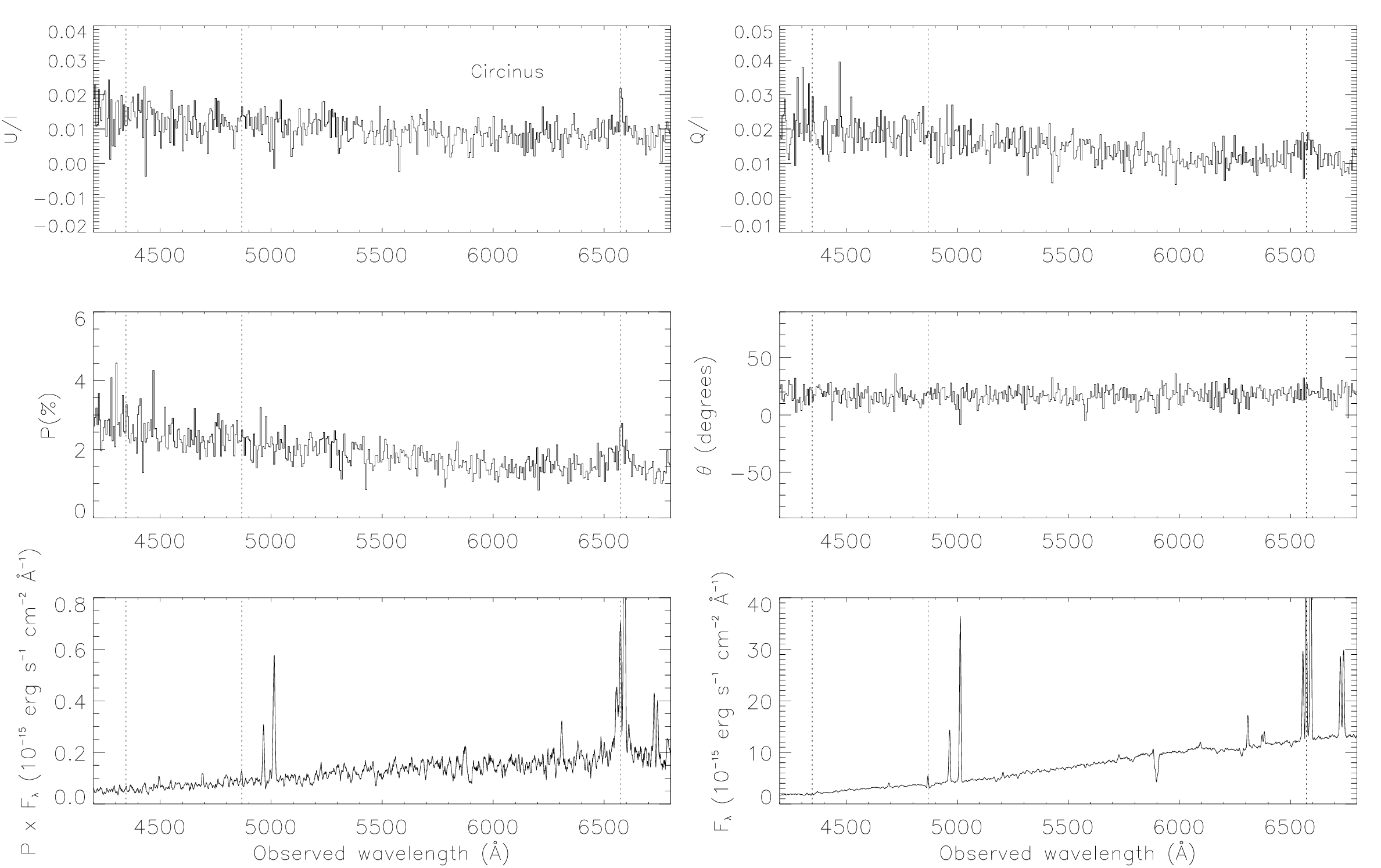}
\caption{Same as in Fig. \ref{fig1}, but for the Circinus galaxy. Note that for this galaxy we have corrected for 
the effect of interstellar polarization, as described in Section \ref{observations}.}\label{A1}
\end{figure*}

\begin{figure*}
\centering
\includegraphics[width=15cm]{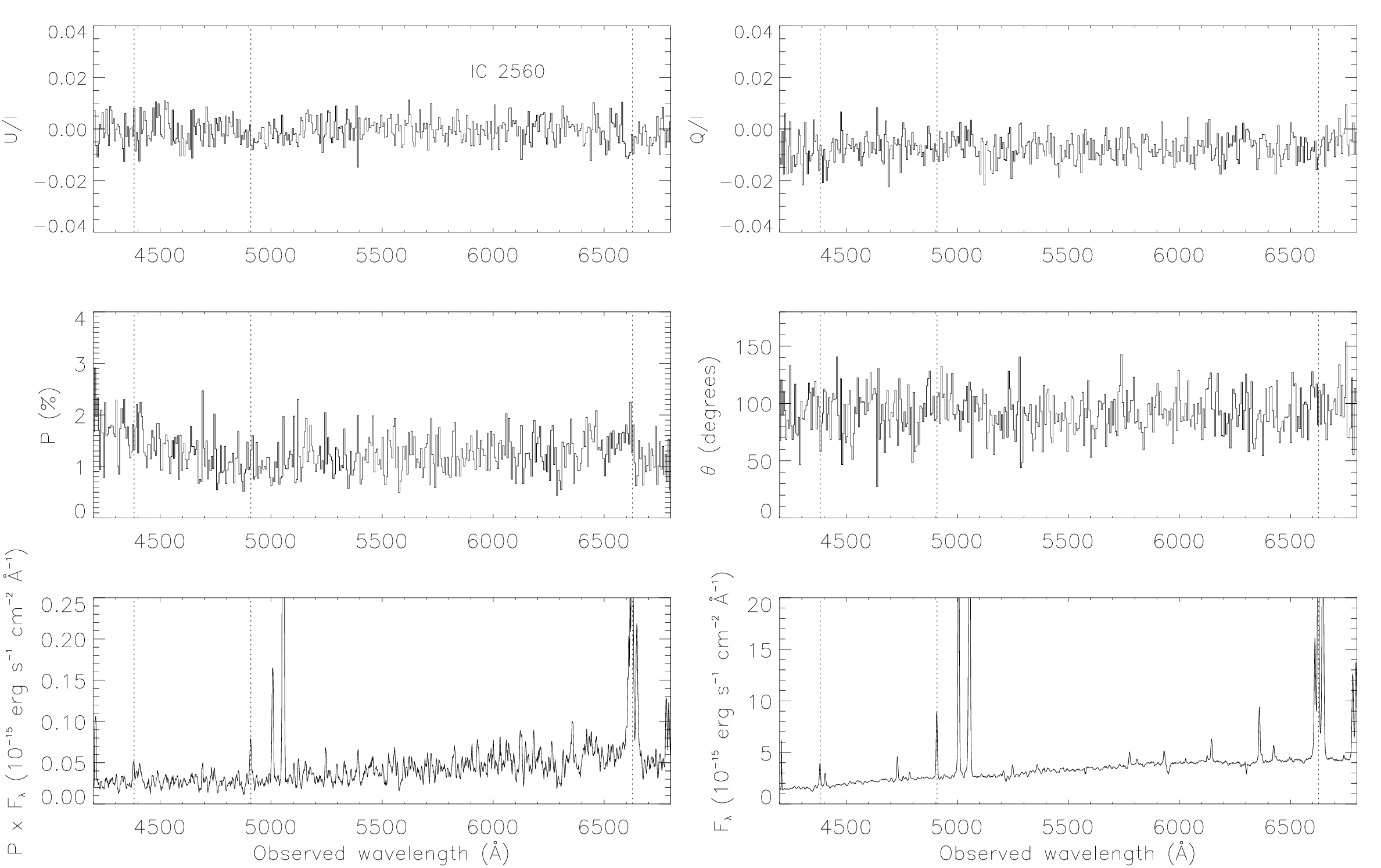}
\caption{Same as in Fig. \ref{fig1}, but for IC\,2560.}\label{A2}
\end{figure*}

\begin{figure*}
\centering
\includegraphics[width=15cm]{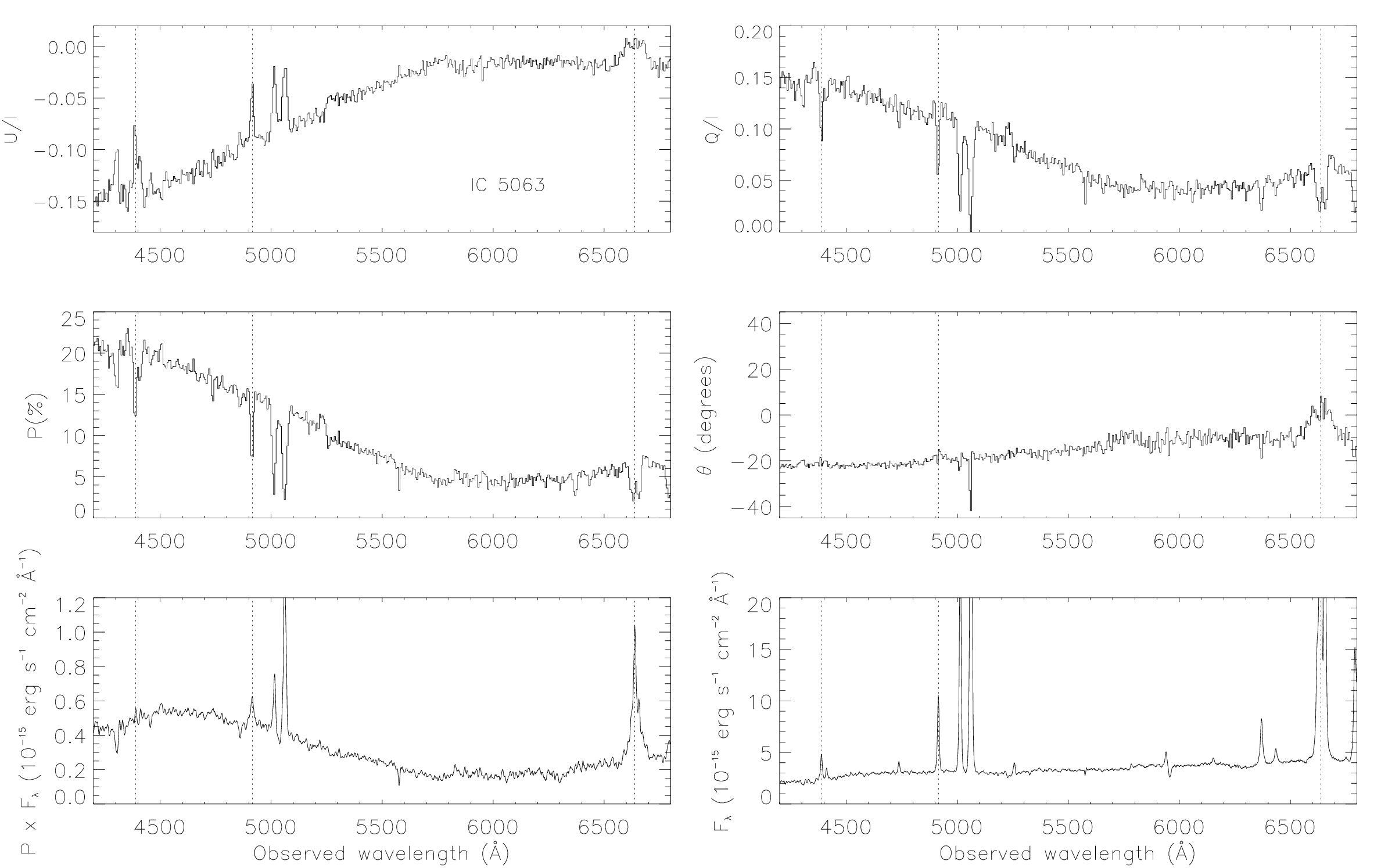}
\caption{Same as in Fig. \ref{fig1}, but for IC\,5063. In the case of this galaxy, which was observed at the 
end of the night, blue wavelengths are affected by daylight polarization, but the broad components of H$\alpha$ and H$\beta$ are
clearly visible,}\label{A3}
\end{figure*}

\begin{figure*}
\centering
\includegraphics[width=15cm]{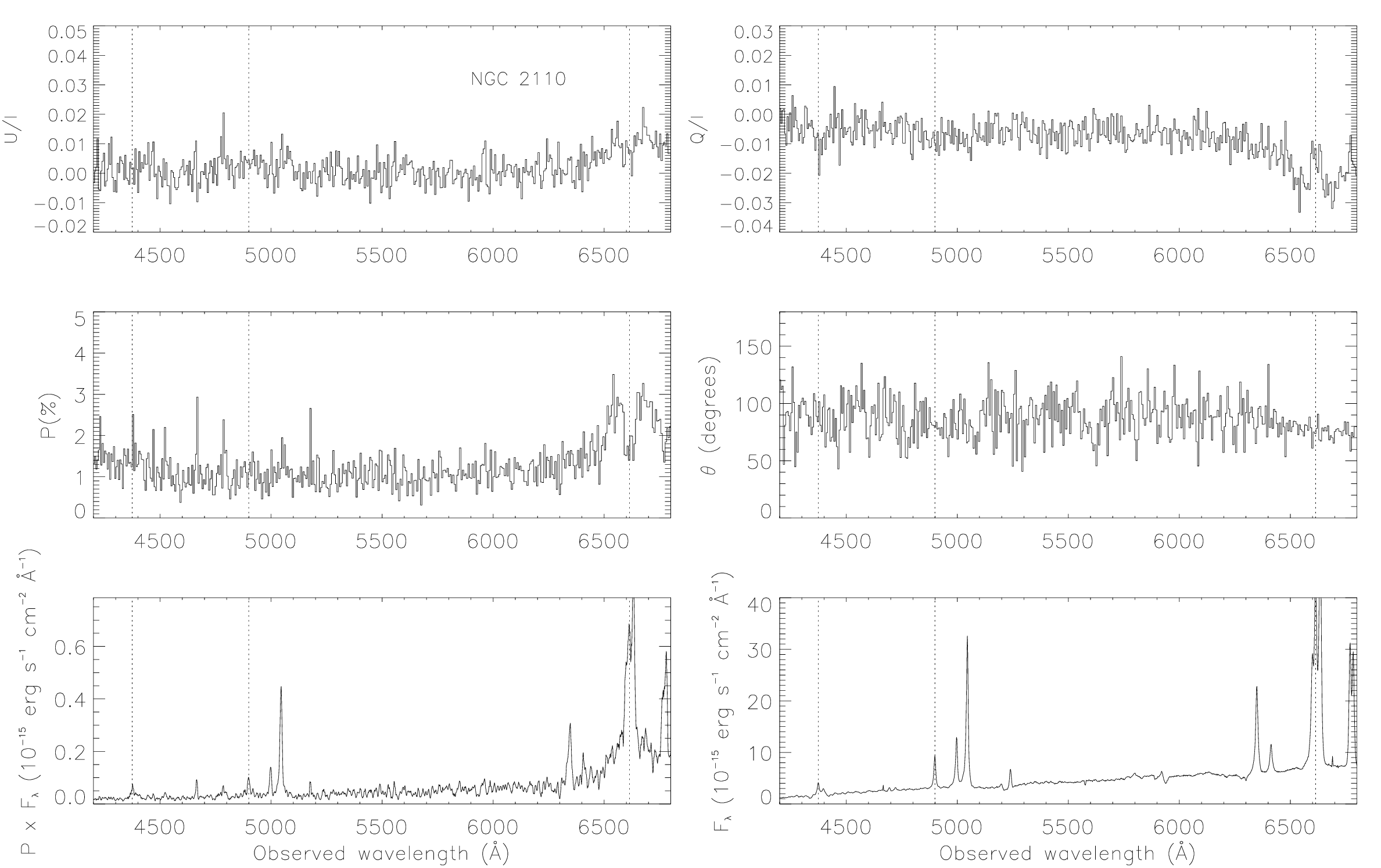}
\caption{Same as in Fig. \ref{fig1}, but for NGC\,2110. Note that for this galaxy we have corrected for 
the effect of interstellar polarization, as described in Section \ref{observations}.}\label{A4}
\end{figure*}

\begin{figure*}
\centering
\includegraphics[width=15cm]{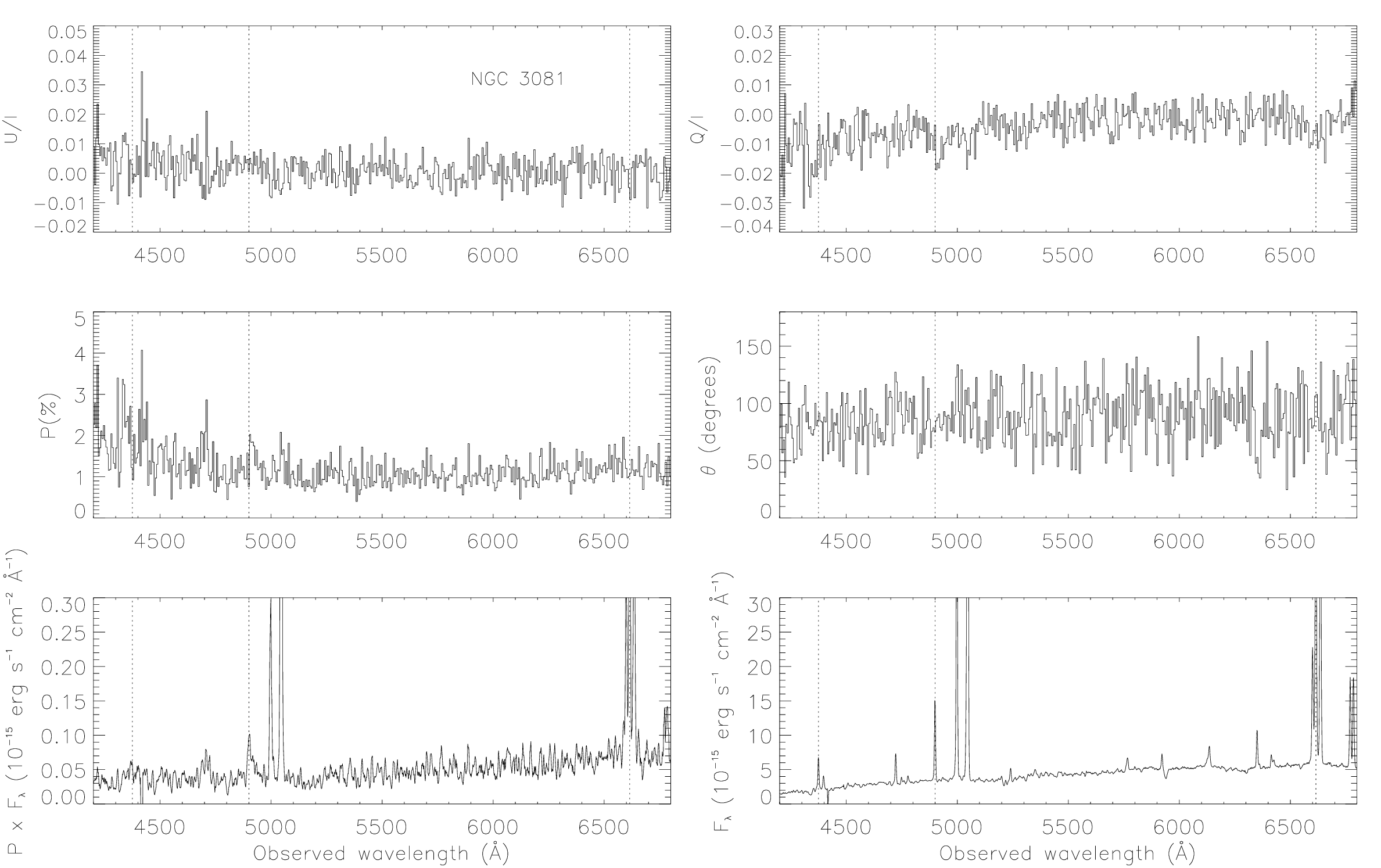}
\caption{Same as in Fig. \ref{fig1}, but for NGC\,3081.}\label{A5}
\end{figure*}

\begin{figure*}
\centering
\includegraphics[width=15cm]{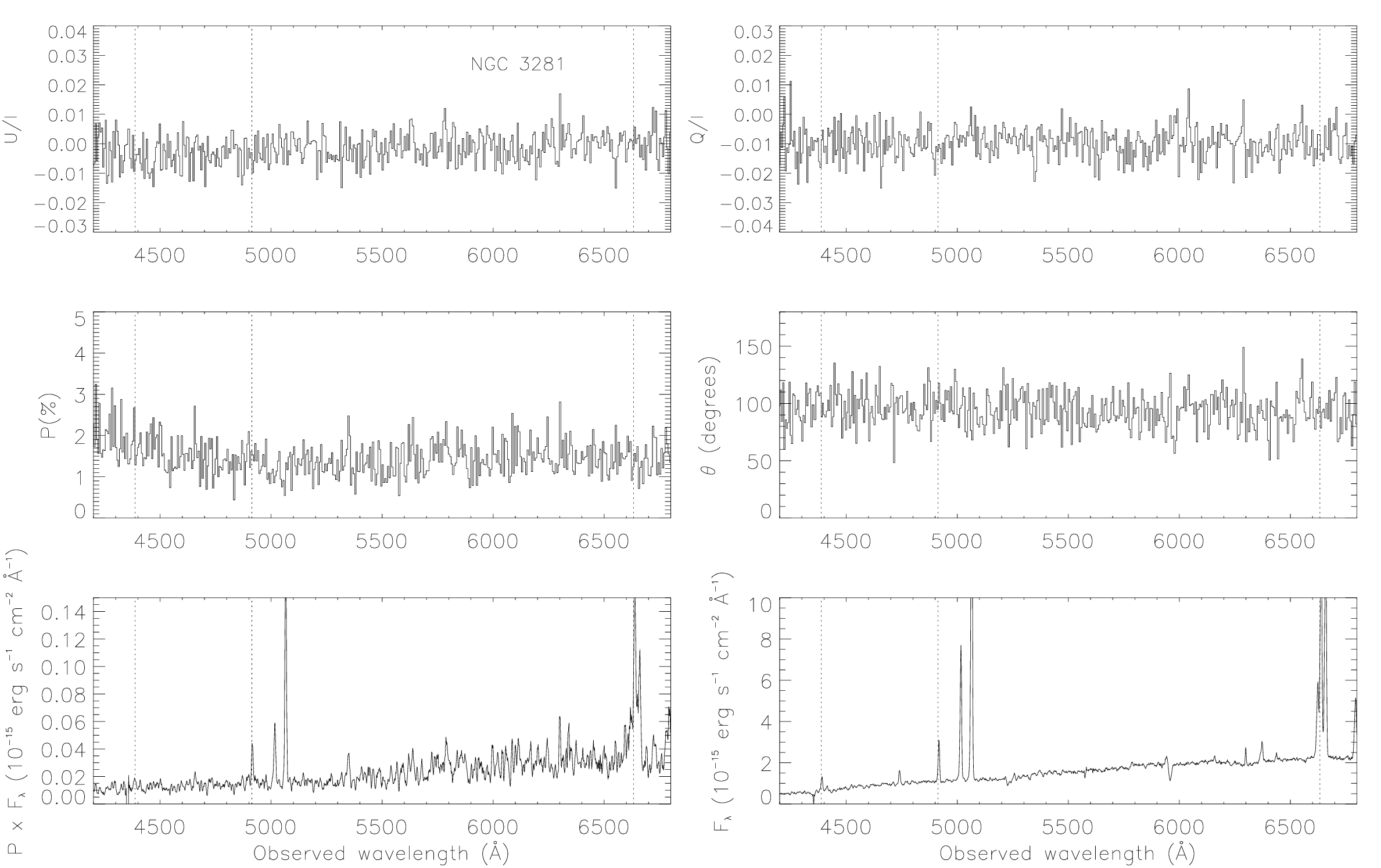}
\caption{Same as in Fig. \ref{fig1}, but for NGC\,3281.}\label{A6}
\end{figure*}

\begin{figure*}
\centering
\includegraphics[width=15cm]{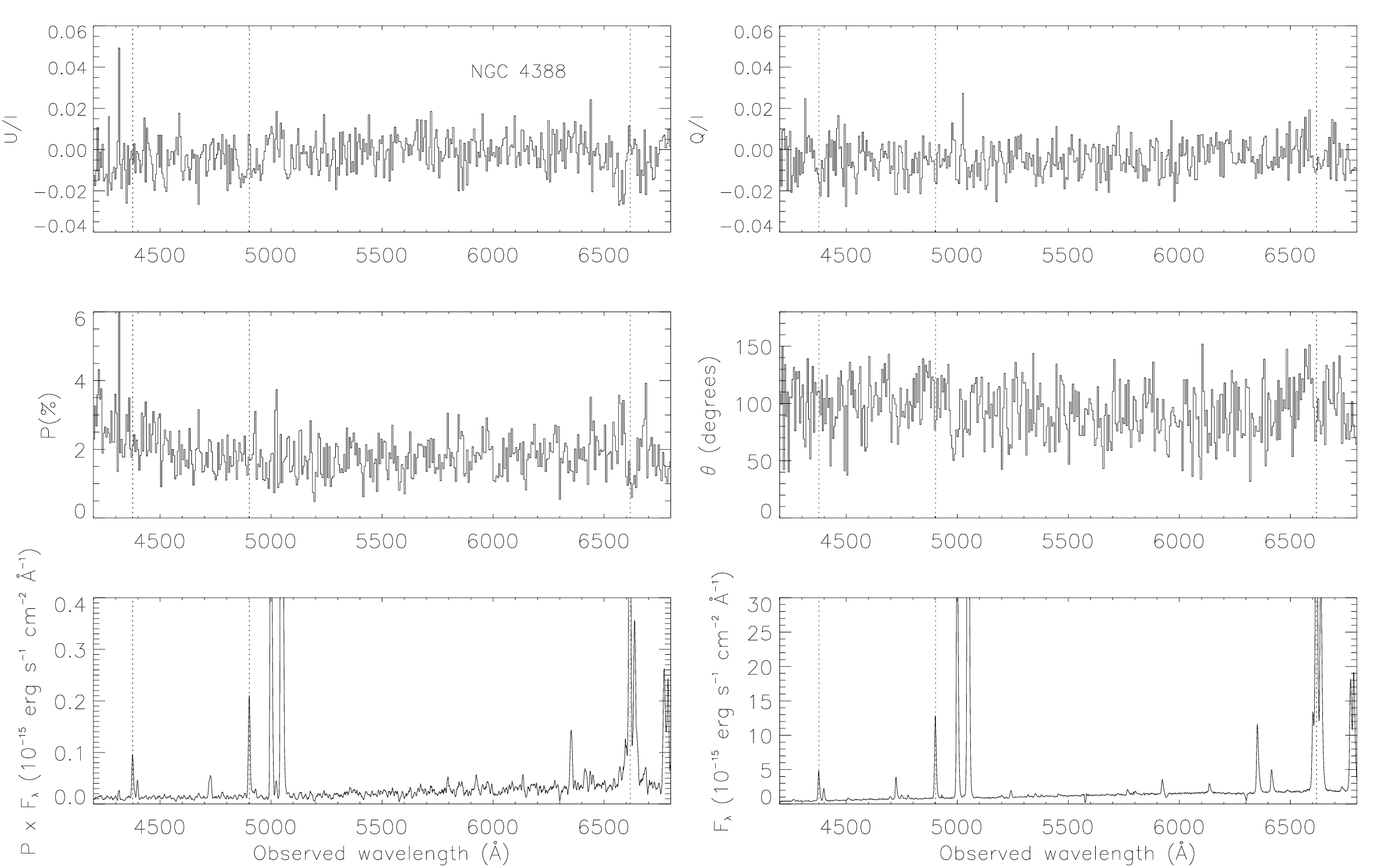}
\caption{Same as in Fig. \ref{fig1}, but for NGC\,4388.}\label{A7}
\end{figure*}

\begin{figure*}
\centering
\includegraphics[width=15cm]{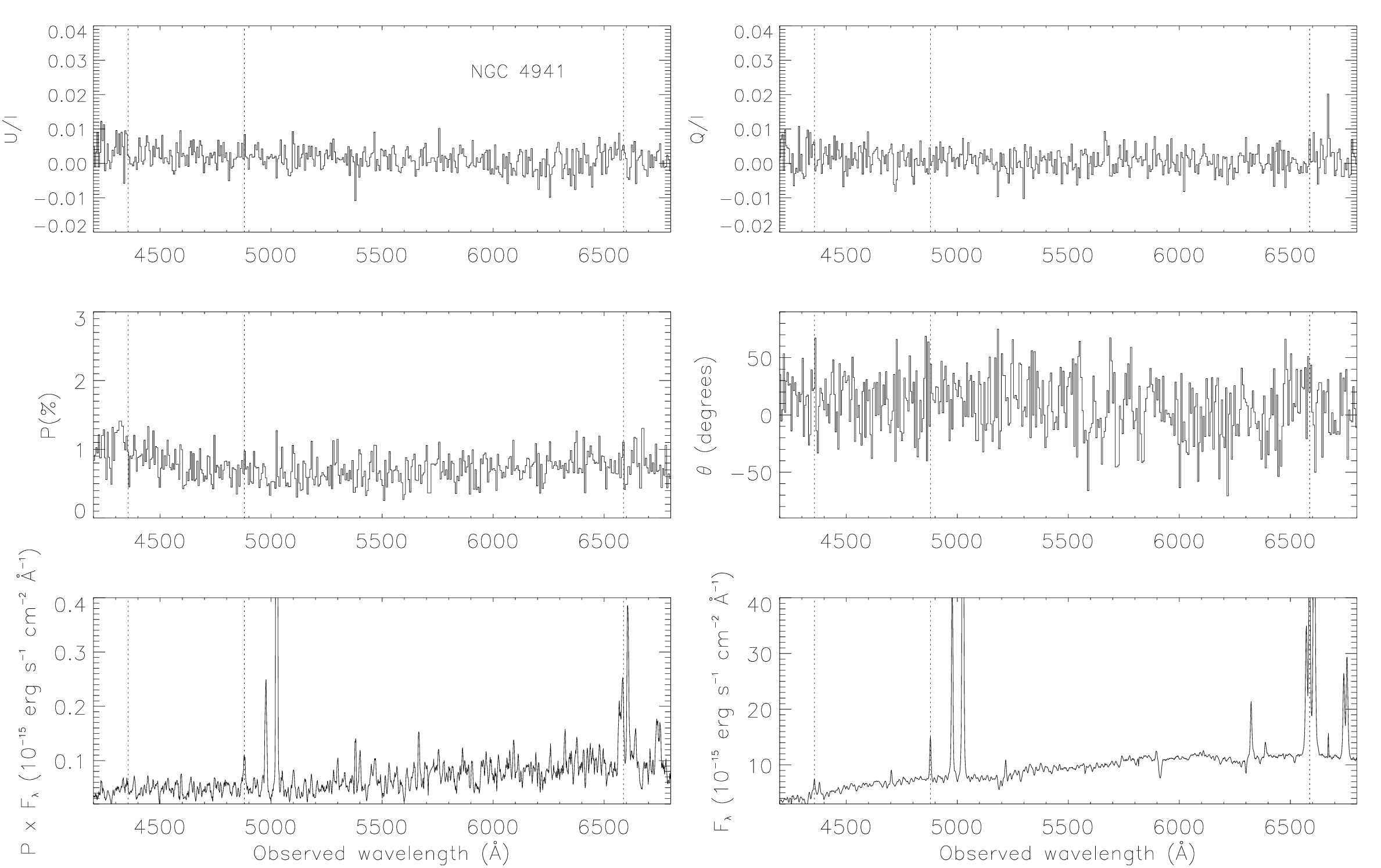}
\caption{Same as in Fig. \ref{fig1}, but for NGC\,4941.}\label{A8}
\end{figure*}

\begin{figure*}
\centering
\includegraphics[width=15cm]{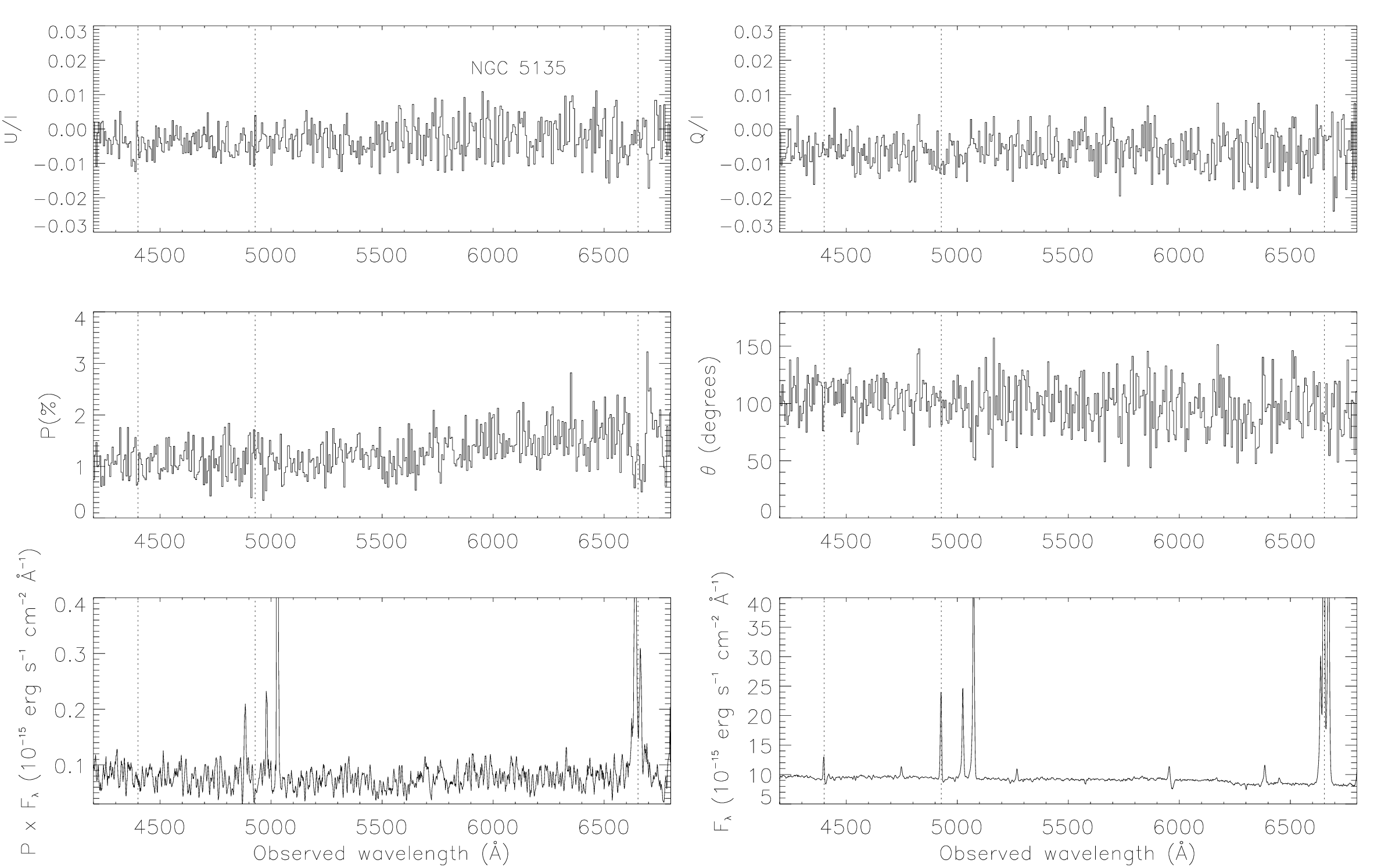}
\caption{Same as in Fig. \ref{fig1}, but for NGC\,5135. Note that for this galaxy, we have applied the corresponding 
starlight dilution-correction using Equation \ref{eq1} and the galaxy fraction spectrum shown in the left panel of 
Figure \ref{fig0}.}\label{A9}
\end{figure*}

\begin{figure*}
\centering
\includegraphics[width=15cm]{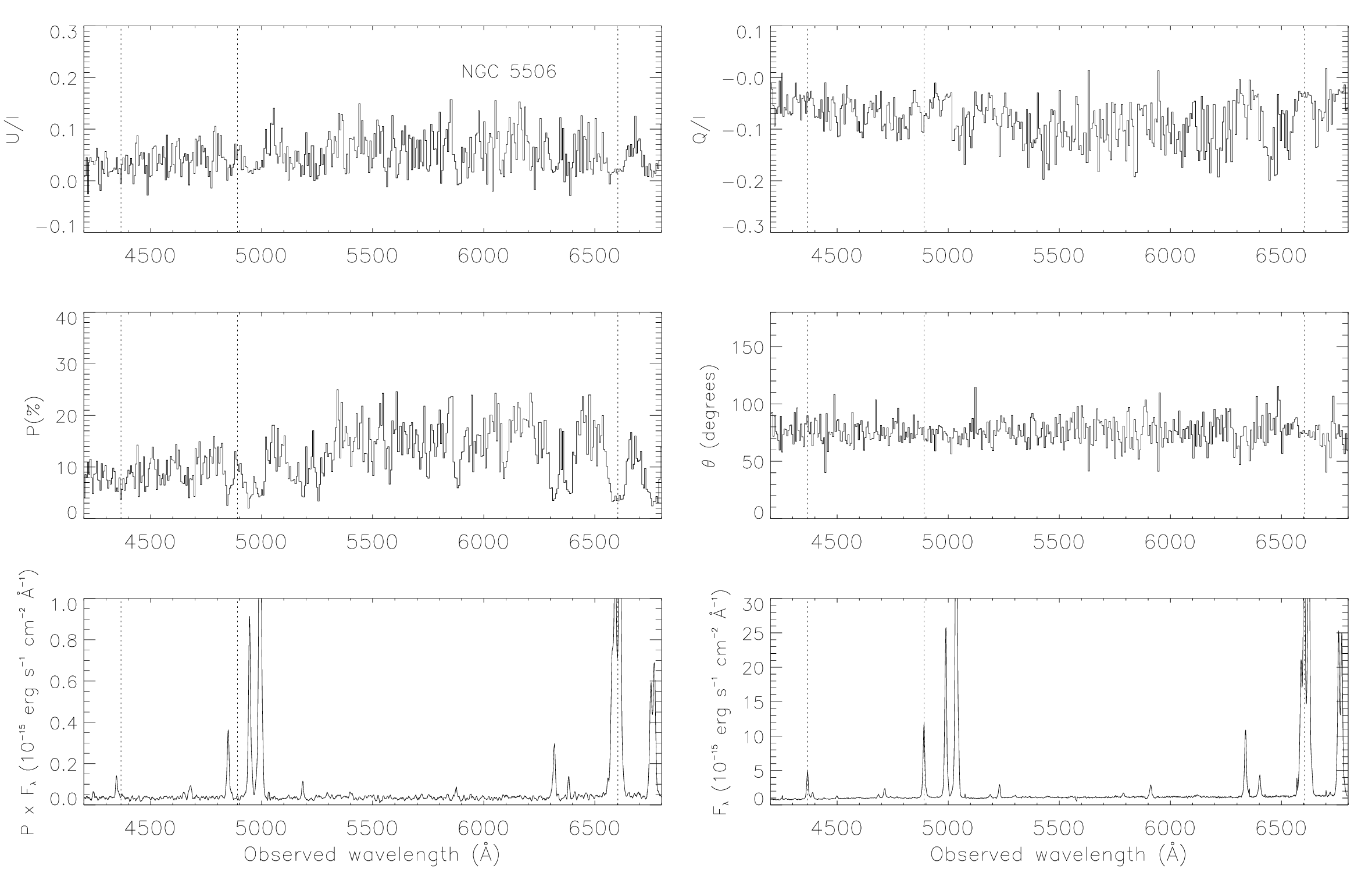}
\caption{Same as in Fig. \ref{fig1}, but for NGC\,5506. Note that for this galaxy, we have applied the corresponding 
starlight dilution-correction using Equation \ref{eq1} and the galaxy fraction spectrum shown in the right panel of 
Figure \ref{fig0}.}\label{A10}
\end{figure*}

\begin{figure*}
\centering
\includegraphics[width=15cm]{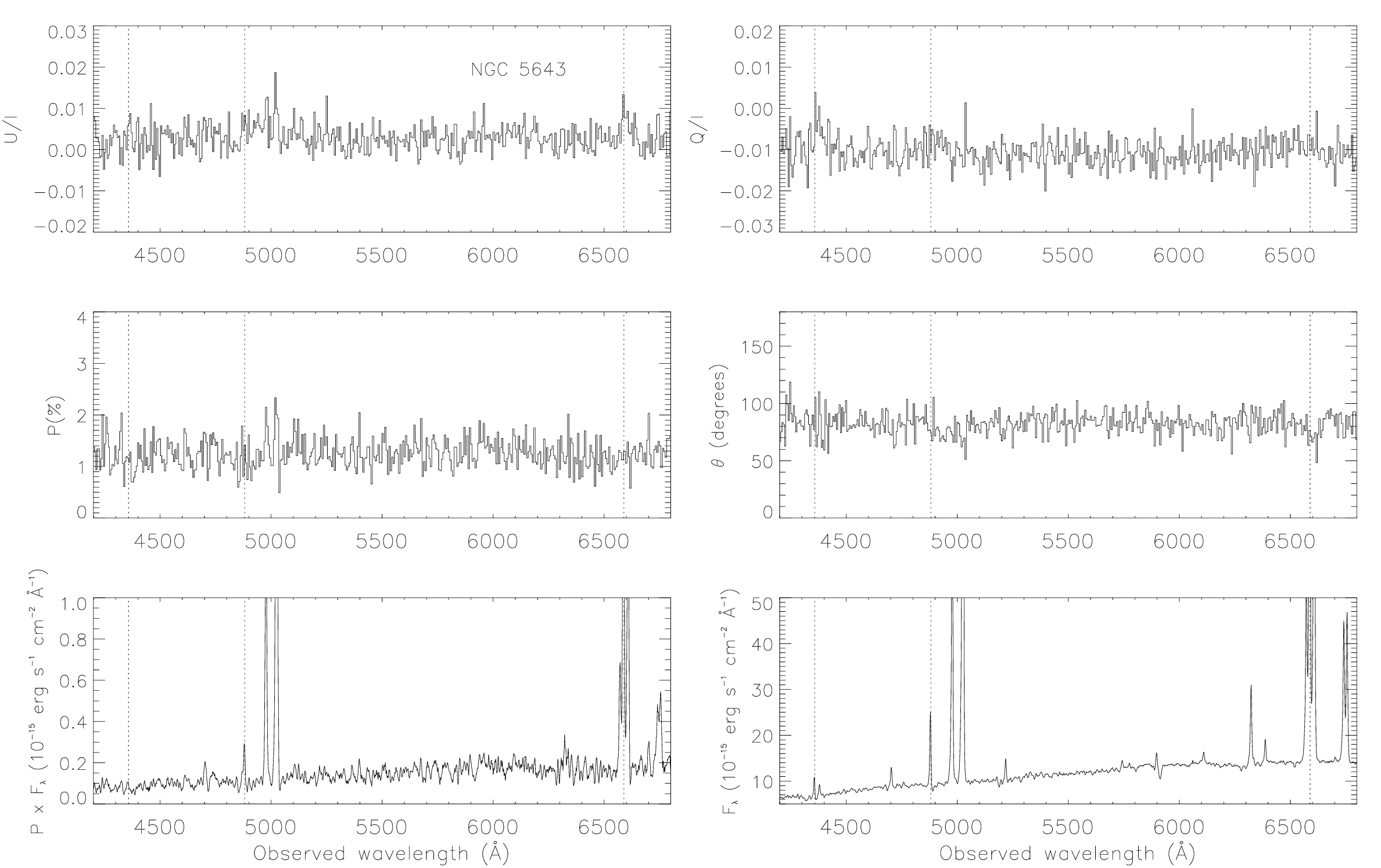}
\caption{Same as in Fig. \ref{fig1}, but for NGC\,5643.}\label{A11}
\end{figure*}

\begin{figure*}
\centering
\includegraphics[width=15cm]{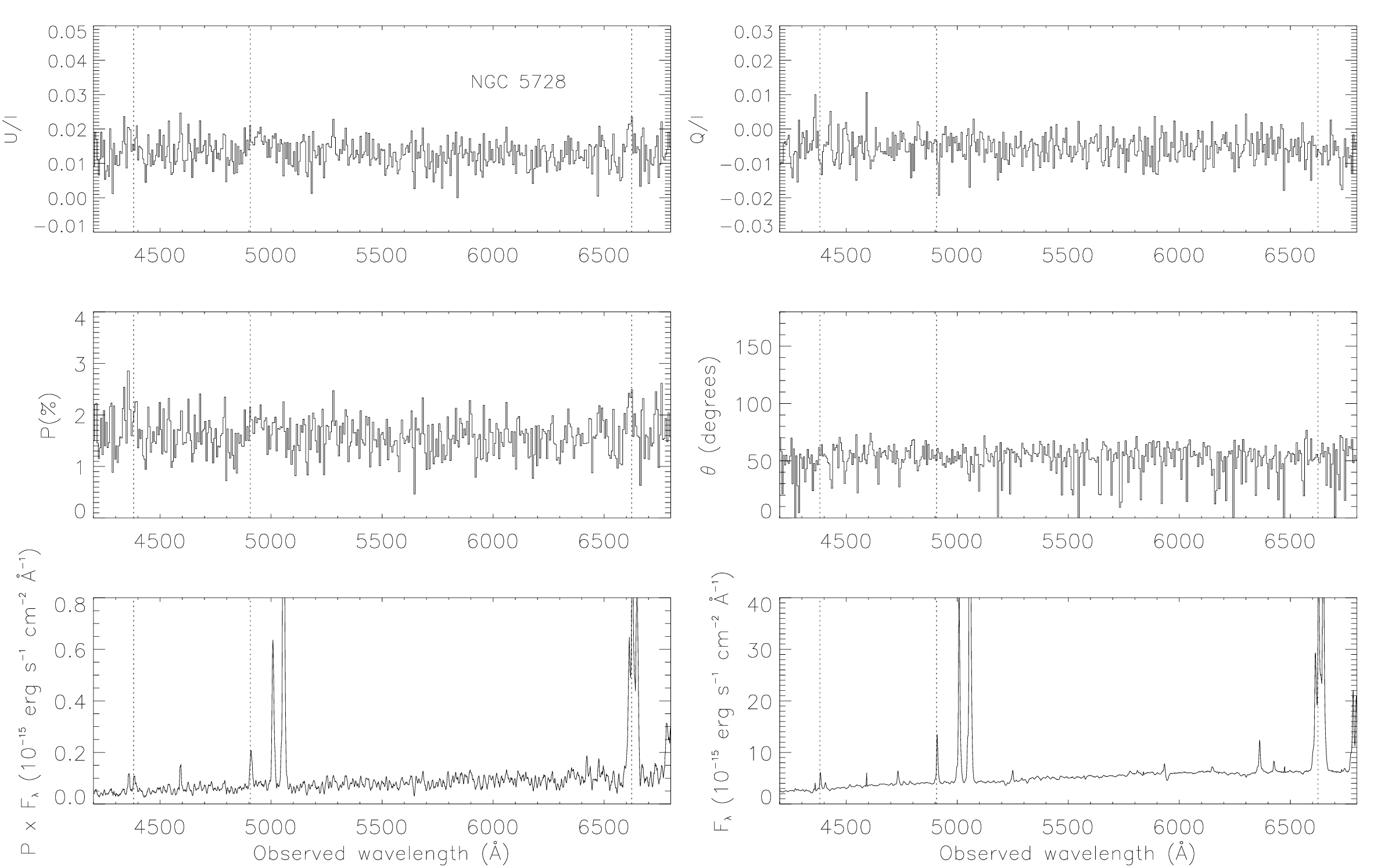}
\caption{Same as in Fig. \ref{fig1}, but for NGC\,5728.}\label{A12}
\end{figure*}

\begin{figure*}
\centering
\includegraphics[width=15cm]{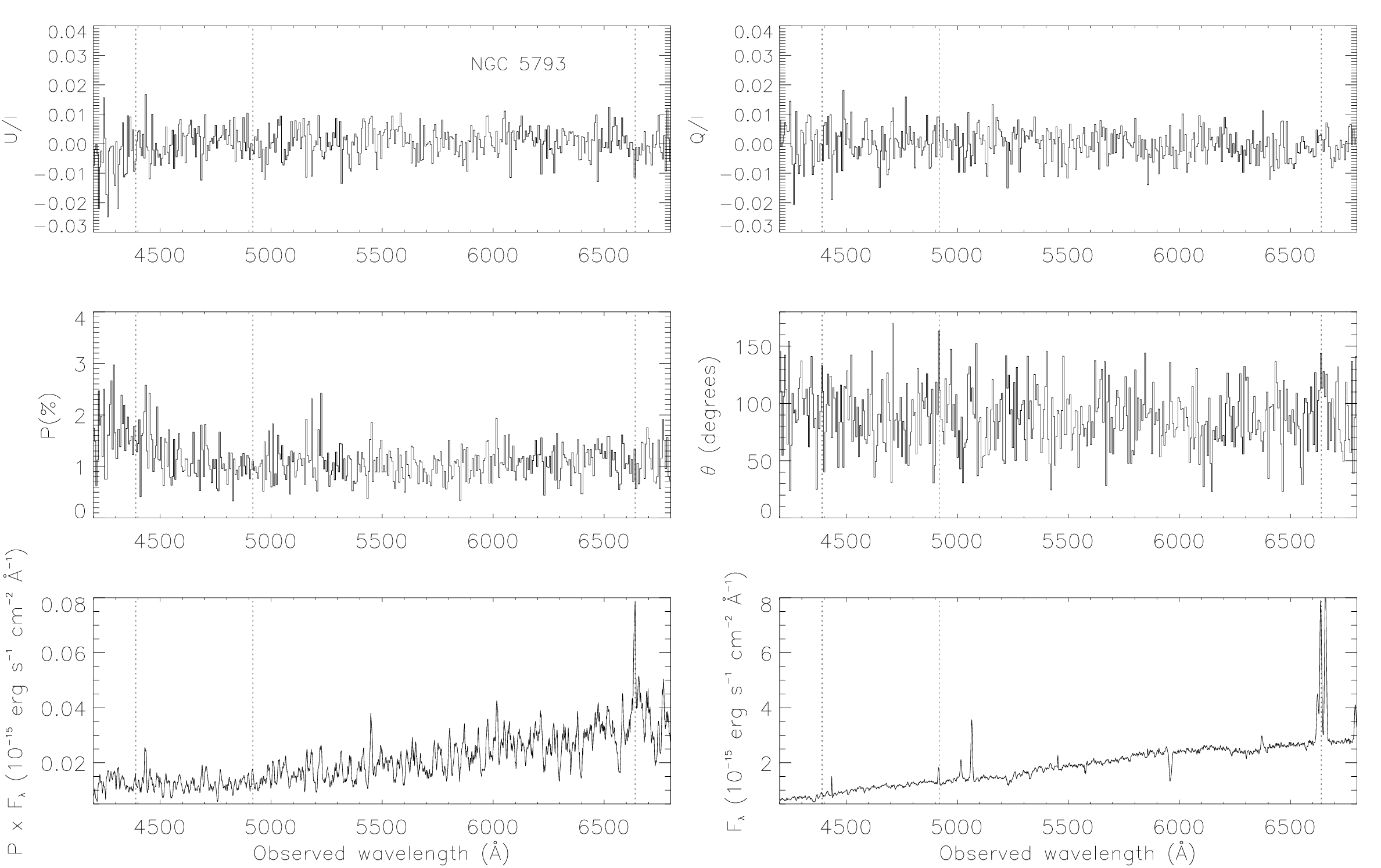}
\caption{Same as in Fig. \ref{fig1}, but for NGC\,5793.}\label{A13}
\end{figure*}

\begin{figure*}
\centering
\includegraphics[width=15cm]{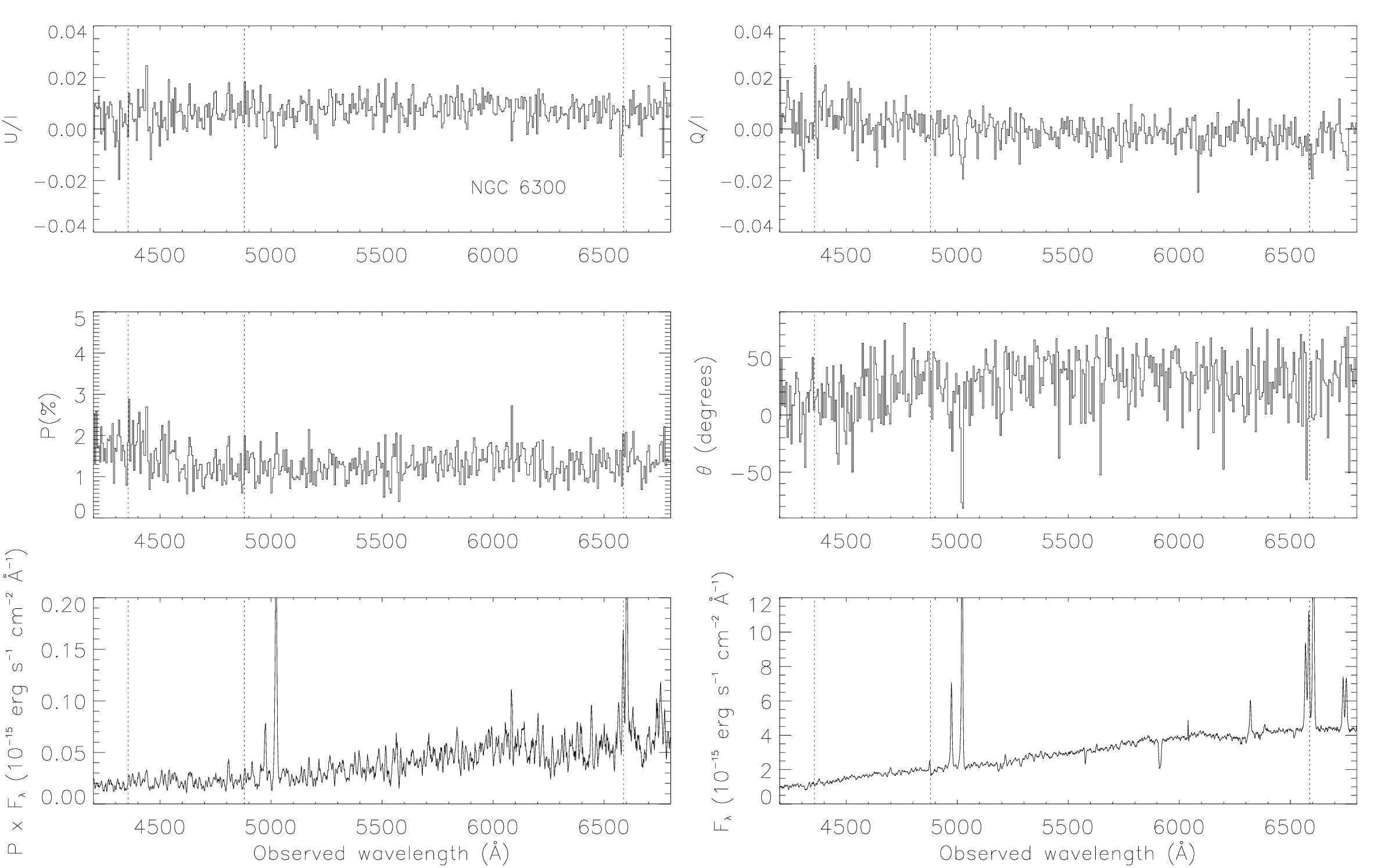}
\caption{Same as in Fig. \ref{fig1}, but for NGC\,6300.}\label{A14}
\end{figure*}

\section{Analysis of \emph{Chandra} and \emph{XMM}-Newton data of NGC\,5793.}
\label{appendixB}

We retrieved from the Heasarc\footnote{http://heasarc.gsfc.nasa.gov} 
archive \emph{Chandra} (ObsIDs 10390) and \emph{XMM}-Newton (ObsID 306050901)
observations of NGC\,5793 with the aim of performing spectral fitting. 
After reducing the two observations, we grouped both \emph{Chandra} and \emph{XMM}-Newton 
spectra to get at least 
15 counts per spectral bin using the {\sc grppha} task. 
The \emph{XMM}-Newton spectrum shows a higher S/N than \emph{Chandra} spectrum. Thus, 
the results presented here are based on the analysis of the \emph{XMM}-Newton spectrum. However, 
we note that we have independently examined the \emph{Chandra} data and all the
results presented here are consistent with them.

%The \emph{Chandra} data were reduced with the {\sc ciao} 4.6 data analysis system 
%and the \emph{Chandra} Calibration Database ({\sc caldb} 4.6.3). The observation 
%was processed to exclude background flares, using the task lc$_-$clean.sl in source-free 
%sky regions of the same observation. The net exposure time after flare removal is 
%31.6 ksec, and the net count rate in the 0.5--10 keV band is 
%$\rm{(6.8 \pm 0.5) \times 10^{-3}~counts~s^{-1}}$. The source spectrum was 
%extracted in a circular region with 2\arcsec~radius centred in the galaxy nucleus. The background 
%spectrum was extracted using a circular region of 18\arcsec~near the source. 
%Response and ancillary response files were created using the CIAO {\sc mkacisrmf} 
%and {\sc mkwarf} tools. 

The \emph{XMM}-Newton observation was reduced using the Science Analysis Software 
(SAS), version 14.0.0. Before the extraction of the spectra, good-timing periods were 
selected (i.e. ``flares'' were excluded). The method used for this purpose maximises the 
S/N ratio of the net source spectrum by applying a different constant count rate threshold 
on the single-events, E $\rm{>}$ 10 keV field-of-view background light curve. The net 
exposure time after flare removal is 11.5 ksec, and the net count rate in the 0.5--10 keV 
band is $\rm{(2.5 \pm 0.1) \times 10^{-3}~counts~s^{-1}}$. The nuclear positions were taken 
from NED, while the extraction region was determined through circles of 20\arcsec~radius 
and the background was selected as a circular region with radius of 40\arcsec~in the same 
chip, free of other sources and as close as possible to the nucleus. The extraction of the source 
and the background regions were done by using the {\sc evselect} task. Response matrix files 
(RMF) were generated using the {\sc rmfgen} task, and the ancillary response files (ARF) were 
generated using the {\sc arfgen} task. 

We performed spectral fitting using the package Xspec within the Heasoft software. 
A reasonable fit is obtained with a combination a non-absorbed power-law 
($\rm{\Gamma=2.0_{-0.5}^{+0.9}}$) and a thermal component (MEKAL, kT$\rm{=0.4_{-0.1}^{+0.1}}$ keV).
We have noticed an excess above $\rm{\sim 4}$ keV compared to the fit. This could
be reproduced by an absorbed power-law component or by the addition of the 
FeK$\rm{\alpha}$ line at 6.4 keV. None of the two models are preferred from the other. 
In the case of an absorbed power-law, the spectral index was fixed to the softer 
power-law obtaining an absorption of $\rm{N_{H}> 1.3\times 10^{23}~cm^{-2}}$. 
In the case of adding a narrow ($\rm{\sigma=0.1~keV}$) FeK$\rm{\alpha}$ line, 
its equivalent width is $1.8\pm^{0.3}_{3.6}$ keV. Note that
both scenarios are consistent with the source being Compton-thick. 
The observed 2--10 keV flux of the source is $\rm{F_{X}=(5.5\pm0.7)\times10^{-14}~erg~s~cm^{-2}}$ 
and the intrinsic 2-10 keV luminosity, once corrected for a factor 70 following \citet{Marinucci12}, 
$\rm{L_{X}=(1.2\pm0.3)\times10^{42}~erg~s^{-1}}$.

Further indication of the Compton-thickness of the source is the ratio between
the X-ray and [O III]$\lambda$5007 \AA~fluxes. From the FORS2 spectra analyzed here we derived 
%an observed [O III] flux of 
%6.4$\times 10^{-15}~erg~s~cm^{-2}$ and 
a reddening corrected [O III] flux of 1.7$\times10^{-12}~erg~s~cm^{-2}$, which gives
%($\rm{H_{\alpha}/H_{\beta}=16.6}$). 
%The final X-ray to [O III] flux ratio is 
$\rm{Log(F_{X}/F_{[O III]})=-1.48}$. Finally, we can also use the
IRAS 12$\rm{\mu m}$ flux ($\rm{F_{12\mu m}}$=160 mJy) to derive the
X-ray to mid-infrared ratio $\rm{log(F_{X}/F_{MIR})=-2.86}$. 
All these results fully support the Compton-thickness nature of NGC\,5793. 

\section*{Acknowledgments}

This research was supported by a Marie Curie Intra European Fellowship within the 
7th European Community Framework Programme 
(PIEF-GA-2012-327934). Based on observations made with ESO Telescopes at the Paranal 
Observatory under programme ID 091.B-0190(A).
MJMG and AAR acknowledge financial support by the Spanish Ministry of Economy and 
Competitiveness through projects AYA2010-18029 (Solar Magnetism and Astrophysical 
Spectropolarimetry) and Consolider-Ingenio 2010 CSD2009-00038. CRA and AAR also 
acknowledge financial support through the Ram\'on y Cajal fellowship. A.A.-H. acknowledges support from AYA2012-31447.
The authors acknowledge support from the COST Action MP1104 ``Polarization as a tool to study the Solar System and beyond'' and
the data analysis facilities provided by the Starlink Project which is run by CCLRC 
on behalf of PPARC. We finally acknowledge R. Antonucci and the anonymous referee for useful comments 
that have substantially contributed to improve this work.

\label{lastpage}

\end{document}